\documentclass[11pt]{article}

\usepackage{cite}
\usepackage{amsmath}
\usepackage{amssymb}
\usepackage{latexsym}
\usepackage{graphicx}

\newcommand{\be}{\begin{eqnarray}}
\newcommand{\ee}{\end{eqnarray}}
\newcommand{\half}{\frac{1}{2}}
\renewcommand{\d}{\partial}
\newcommand{\nn}{\nonumber}
\newcommand{\bea}[1]{\left( \begin{array}{#1}}
\newcommand{\ena}{\end{array} \right)}

\newcommand{\CA}{{\cal A}}

\newcommand{\CC}{{\cal C}}

\newcommand{\CM}{{\cal M}}

\newcommand{\CO}{{\cal O}}

\renewcommand\bar{\overline}

\renewcommand\>{\rangle}
\newcommand\<{\langle}

\newcommand\w\omega

\newcommand\z\zeta
\renewcommand\l\lambda
\newcommand\f\phi
\newcommand\De\Delta
\newcommand\s\sigma
\newcommand\oo\infty
\newcommand\g\gamma

\newcommand\e\epsilon
\renewcommand\a\alpha
\renewcommand\b\beta
\newcommand\de\delta

\renewcommand\th\theta

\renewcommand{\tilde}{\widetilde}
\renewcommand{\hat}{\widehat}

\newcommand{\sect}[1]{\section{#1}\setcounter{equation}{0}}

\newcommand\lrpar{\raise .8ex\hbox{$^\leftrightarrow$} \hspace{-9pt}
\partial}
\newcommand\lpar{\raise .8ex\hbox{$^\leftarrow$} \hspace{-9pt}
\partial}
\newcommand\rpar{\raise .8ex\hbox{$^\rightarrow$} \hspace{-9pt}
\partial}

\newcommand{\gsim}{\lower.7ex\hbox{$\;\stackrel{\textstyle>}{\sim}\;$}}
\newcommand{\lsim}{\lower.7ex\hbox{$\;\stackrel{\textstyle<}{\sim}\;$}}
\begin{document}

\baselineskip=18pt

\setcounter{footnote}{0}
\setcounter{figure}{0}
\setcounter{table}{0}

\begin{titlepage}

\begin{center}
\vspace{1cm}

{\Large \bf  Scattering States in AdS/CFT}

\vspace{0.8cm}

{\bf A. Liam Fitzpatrick$^1$, Jared Kaplan$^2$}

\vspace{.5cm}

{\it $^1$ Department of Physics, Boston University, \\ Boston, MA 02215, USA}

\normalsize\emph{ $^2$  SLAC National Accelerator Laboratory,  \\ 2575 Sand Hill Road, Menlo Park, CA 94025, USA}

\end{center}
\vspace{1cm}

\begin{abstract}

We show that suitably regulated multi-trace primary states in large $N$ CFTs behave like `in' and `out' scattering states in the flat-space limit of AdS.  Their transition matrix elements approach the exact scattering amplitudes for the bulk theory, providing a natural CFT definition of the flat space S-Matrix.  We study corrections resulting from the AdS curvature and particle propagation far from the center of AdS, and show that AdS simply provides an IR regulator that disappears in the flat space limit.

\end{abstract}

\bigskip
\bigskip

\end{titlepage}

\sect{Introduction}

We have gained enormous insights from the holographic description of quantum gravity provided by the AdS/CFT correspondence\cite{Maldacena:1997re,Witten:1998qj,Gubser:1998bc}, so searching for a holographic description of flat space is a natural next step.  We would expect a dual theory to compute the S-Matrix, since this is the only exact observable of quantum gravity in flat space.  One approach to this problem is to study the perturbation expansion of the S-Matrix, in which remarkable patterns have been found, in the hopes of finding a weak-weak dual description \cite{ArkaniHamed:2008gz, ArkaniHamed:2009dn}.  Another approach, which dates back to the early days of AdS/CFT \cite{Polchinski:1999ry,Susskind:1998vk}, is to compute the flat space S-Matrix as the limit of observables in AdS, computed via AdS/CFT.  In this case, we have a weak-strong coupling duality, and the dual description is the limit of a CFT.

Despite very exciting recent progress on this question \cite{mellin,Okuda:2010ym,Gary:2009ae, 
Penedones:2007ns, Cornalba:2007zb} and plausible physical expectations from early work \cite{Polchinski:1999ry,Susskind:1998vk,Giddings:1999jq}, there has been some controversy \cite{Gary:2009mi} as to whether the S-Matrix can be extracted with arbitrary precision from CFT correlators.  We will clarify these issues by providing a precise construction of the CFT states that turn into in and out states  in the flat space limit.  As opposed to working with awkward wavepackets, we will see that scattering states are very natural from the CFT point of view -- they are simply regulated primary states.  We will also calculate infrared effects and show that in the flat space limit we recover the S-Matrix, without any contributions from outside of `our universe' at the center of AdS.

In order to define an S-Matrix one must understand a theory in the infrared and identify the particles.  This means that a theory with an S-Matrix should be describable at low energies by an effective field theory, so we will explicitly use the language of EFT when discussing the AdS dynamics, focusing on CFTs whose AdS duals are controllable EFTs with a flat-space limit.  Such ``effective conformal theories'' \cite{ECFT}
have a cut-off in the dimensions of operators, dual to the energy
cut-off in the AdS EFT, and a sector of low-dimension primary operators
below the cut-off. The flat space limit consists of sending the AdS length scale $R \to \infty$ while keeping local energies and distances fixed.
However, the Planck scale is typically some power of $N$ in the CFT
times the AdS curvature scale $R^{-1}$, so we will need CFTs with a large free
parameter $N$ in order to have a range of energies below the cut-off 
within which AdS looks approximately flat.  
 We will remain agnostic about the presence of extra dimensions and stringy 
degrees of freedom, but we expect that in a UV complete description 
we will have something like string modes, and so we will need to send the string scale, and 
therefore the 't Hooft coupling, to infinity via $\lambda \to \infty$.  Extra 
dimensions of radius $R$ are expected in the bulk description of 
superconformal theories, but since we will not be assuming the presence of 
supersymmetry, we will mostly ignore them.

We should emphasize that although we will use the language of effective field theory and the properties of free fields in AdS to motivate our asymptotic initial and final states, once the S-Matrix has been defined in terms of CFT observables,  it should be valid non-perturbatively, even in regimes where we expect to form sub-AdS scale black holes as intermediate states in scattering processes.  This follows from the usual S-Matrix intuition that no matter how strong the interactions are, they become vanishingly small once the particles become well-separated, and that due to conformal/Lorentz invariance, we can describe arbitrarily boosted states in isolation.  

An elementary but useful tool for understanding the correspondence between 
the states in the CFT$_d$ and in AdS$_{d+1}$ is provided by an explicit identification of 
the Fock spaces of the two descriptions\cite{Balasubramanian:1998de,Witten:1998qj,Balasubramanian:1998sn,Balasubramanian:1999ri,Banks:1998dd}, which we review in section 
\ref{sec:hilbertspaces}. The identification is simplest in the
case of a scalar field in AdS$_{d+1}$, which has a discrete spectrum in global coordinates:
\be
\phi(x) &=& \sum_{n,l,J} \phi_{n,l,J} (x) a_{n,l,J} + \phi^*_{n,l,J} a_{n,l,J}^\dagger,
\ee
where the wavefunctions $\phi_{n,l,J}$ are solutions of the Klein-Gordon equation in $(d+1)$ dimensions.  This Fock space description is the familiar result of canonical quantization.  What is perhaps less  familiar is the fact that the scalar operators $\CO$ dual
to $\phi$ have a parallel mode decomposition in Lorentzian radial quantization:
\be
\CO(x) &=& \sum_{n,l,J} \frac{1}{N^{\CO}_{n,l,J} }
\left( e^{i (\Delta+2n +l)t}Y_{lJ}(\hat{x}) a_{n,l,J} + e^{-i(\Delta+2n+l)t} Y_{lJ}^*(\hat{x}) a_{n,l,J}^\dagger \right), 
\label{eq:CFTFock}
\ee
where here $x$ is a $d$-dimensional position and $\Delta$ is the conformal 
dimension of $\CO$. In the limit of infinite
$N$ in the CFT, the CFT has its own exact Fock space representation
given by the creation/annihilation operators in (\ref{eq:CFTFock}). 
Thus, perturbatively in $1/N$, we can 
use this correspondence to explicitly identify an operator in the CFT that 
will create a particle in AdS with any wavefunction.  This means that we will 
always discuss the physics in terms of well-defined, normalized states, and 
in particular we will avoid all discussion of so-called non-normalizable modes.

Our major application of this state-state correspondence will be to
construct states in the CFT that are appropriate for scattering
in the flat-space limit of AdS.  Physically, it should be clear 
 that these scattering states exist in AdS:  observers in a universe with an AdS radius the size of
Geneva would certainly find the same scattering amplitudes as we are now seeing at the LHC.  
These AdS scattering states necessarily have exact CFT duals, but what is perhaps
surprising is that the center-of-mass in and out scattering states 
are not exotic from the CFT point of view, but rather
they are simply regulated multi-trace primary states with very large dimension.
We want to be able to use CFT correlation functions to calculate 
bulk S-matrix elements, so we will show how
one can integrate products of CFT primary operators against
certain smearing functions in order to pick out the appropriate
scattering states with $n$ particles:
\be
\mathcal{C}_f = \int \prod_{i=1}^n \left[ d^d x_i R\left(\frac{t_i}{\Delta T} \right) \right] f(x_1,x_2,...,x_m) \CO(x_1)... \CO(x_m) ,
\ee
The regulator function $R$ limits the region of time integration to a window $\Delta T \ll R$, so that $\mathcal{C}_f$ prepares states consisting of well-separated particles near the boundary of AdS.
This is similar to wavefunctions $F$ that one typically integrates
against flat space fields $\phi$ in order to extract in and out
states when deriving the LSZ formula, as we review in appendix \ref{sec:AppendixLSZ}, where $\phi_F \sim \int d^d x (\phi \dot{F}^* - F \dot{\phi}^*)$. The major difference is that if one is restricted to work
with boundary operators, then one must smear over time as well as
spatial coordinates.  Once we have constructed the operators $\mathcal{C}_f$ and the corresponding normalized states, scattering amplitudes are simply the matrix elements
\be
S_{g f} =  \< 0 | \mathcal{C}_g^\dag \ \! \mathcal{C}_f | 0 \>
\ee
where the $f$ and $g$ states are separated in time by $\pi R$, so that they represent in and out states, respectively.  We will solve explicitly for the functions $f$ in AdS$_3$ and check them in a large class of examples by using them
to extract S-matrix elements from four-point correlation functions of massless scalar fields interacting through
contact terms in AdS, taking advantage of some clever observations from \cite{Gary:2009ae}.  

Finally, we will discuss at length the issue of the decoupling of long-distance
curvature effects in the flat space limit of AdS.  For the purposes of understanding the flat space limit and the S-Matrix, it is best to think of AdS as a very symmetrical and intelligent box.   As shown in \cite{Callan:1989em}, AdS acts naturally as an IR regulator because a space-like sphere of radius $\kappa$ has a surface area that grows as $e^{\kappa/R}$, so that by Gauss's law, even massless fields fall off exponentially with distance.  This means that once particles exit the approximately flat universe of size $R$ in the center of AdS, they propagate without interactions.  The intuitive fact that there is exponentially more volume near the boundary of AdS as compared to flat space simply means that particles quickly become well-separated.  A related  but rather formal AdS nicety is that massless particles can be given a precise definition, in terms of their unique lowest energy state.  Thus flat-space IR divergences arising from massless particle
exchange or emission become finite in AdS, and in many ways AdS amplitudes are actually
much better-behaved than those of Minkowski space.  One can compute IR safe observables by using the AdS radius $R$ as an infrared regulator, and then take $R$ to infinity to obtain flat space rates and cross sections.

The outline of the paper is as follows.
In section 2, we discuss the correspondence between the Fock spaces 
of AdS and its CFT dual, and use this to construct in and out 
states from CFT primaries.  Along the way, we show how the conformal
algebra reduces to the Poincar\'e symmetry in the bulk when acting
on high dimension operators.  In section 3, we show that our
construction for in and out states reproduces the correct S-matrix amplitudes
for massless particles in AdS$_3$ interacting through contact terms.
In section 4, we discuss the effects of IR divergences in the presence
of AdS curvature, and how AdS acts as an IR regulator.  In section 5
we discuss future directions and conclude.  Throughout we will work
in units where the AdS length $R = 1$ when it is not explicitly specified.

\sect{Flat Space Scattering States from a CFT}
\label{sec:States}

In this section we will identify the sub-space of CFT states that correspond to in and out states in the $R \to \infty$ limit, and understand how the conformal algebra reduces to the Poincare algebra when acting on these flat-space states.   A direct equivalence between the Fock space of free particles in AdS and states in the boundary CFT will prove useful for understanding how the CFT can be used to setup a bulk scattering process.   We will see that multi-trace primary states form a particularly natural and well-defined set of in and out states from the CFT perspective.  We also construct plane waves, which are more familiar from a bulk point of view, but perhaps less natural as CFT states.  Our setup generalizes to all orders in perturbation theory and even into the non-perturbative regime, with the caveat that as in flat space, it is somewhat non-trivial to define the notion of a single particle in a non-perturbative way.  In appendix \ref{sec:AppendixLSZ} we explain the relationship between our construction and the LSZ prescription in flat spacetime.

\subsection{Dual Hilbert Spaces}
\label{sec:hilbertspaces}

The AdS/CFT correspondence is most often described as a dictionary between the AdS path integral and the generating function of CFT correlators.  However, AdS/CFT also implies an isomorphism between the Hilbert spaces of the bulk and boundary theories.  We will begin with an explicit realization of this isomorphism, allowing us to develop a precise bulk-boundary intuition.  Consider AdS in global coordinates, with Lorentzian metric
\be
ds^2 = \frac{1}{\cos ^2  (\rho)} \left(-dt^2 + d \rho^2 + \sin^2 (\rho) d \Omega^2 \right) ,
\ee
where the boundary corresponds to $\rho = \pi/2$, and has topology $R \times S^{d-1}$.  Via AdS/CFT, the global time coordinate $t$ in AdS relates to the scale in the CFT -- not to the usual Minkowski `time' -- so that $t$ translations in the CFT are generated by the dilatation operator.  In the more commonly encountered Euclidean version, $it$ becomes the radius in radial quantization, but we will work with Lorentzian coordinates because we want to study scattering.

If we canonically quantize a free scalar field on an AdS background, we find
\be
\phi(t, \rho, \hat{x}) = \sum_{n,l,J} \phi_{nlJ}(t, \hat{x},  \rho) a_{nlJ} + \phi_{nlJ}^*(t, \hat{x},  \rho) a_{nlJ}^\dag ,
\ee
where the wavefunctions $\phi_{nlJ}(x)$ take the form
\be
\label{eqn:phinlj}
\phi_{nlJ}(t, \hat{x},  \rho) = \frac{1}{N_{\Delta nl}^\phi} e^{i E_{n,l} t} Y_{lJ} (\hat{x}) \left[ \sin^l \rho \cos^\Delta \rho F \left(-n, \Delta+l+n, l+\frac{d}{2}, \sin^2 \rho \right) \right] ,
\ee
with $E_{nl} = \Delta + 2n + l$ and $m^2 = \Delta(\Delta - d)$,  $Y_{lJ}$ are the normalized spherical harmonics, and $F =_{\ 2} \! \! F_1$ is the Gauss hypergeometric function.  The normalization factor $N_{\Delta nl}^\phi$ is fixed by the canonical commutation relations, giving
\be
\label{eqn:phiflat}
N_{\Delta nl}^\phi = 
(-1)^n \sqrt{\frac{n! \Gamma^2(l + \frac{d}{2}) \Gamma(\Delta + n - \frac{d-2}{2})} {\Gamma(n+l+\frac{d}{2}) \Gamma(\Delta +n+l)} } .
\ee
If we consider these wavefunctions $\phi_{nlJ}$ in the limit that $\rho \ll 1$, which means that we are in a region much smaller than the AdS curvature scale $R$, then they approach the spherical Bessel functions that appear when we quantize a field in flat spacetime.  Specifically, one finds that
\be
\frac{1}{N_{\Delta nl}^\phi} \sin^l \rho \cos^\Delta \rho F \left(-n, \Delta+l+n, l+\frac{d}{2}, \sin^2 \rho \right)  
\to \frac{1}{\rho^{\frac{d-2}{2}}} J_{l+(d-2)/2} \left( \left(E_{n,l}^2-\Delta^2 \right)^{1/2} \rho \right)
\ee
This will be very useful later on, when we make contact with flat space and scattering amplitudes.

AdS/CFT implies that there exists an operator $\CO$ in the dual CFT that can be quantized in terms of these same creation and annihilation operators
\be
\label{eqn:OQuant}
\CO(t, \hat{x}) = \sum_{n,l,J} \frac{1}{N_{nlJ}^\CO}  \left( e^{i E_{n,l} t} Y_{lJ} (\hat{x}) a_{nlJ} +   e^{-i E_{n,l} t} Y_{lJ}^* (\hat{x})a_{nlJ}^\dag \right) .
\ee
The normalizations $N_{\Delta n l}^\CO$ are fixed by requiring that the two-point function $\langle \CO(x) \CO(y) \rangle$ takes the correct form, as we show explicitly in Appendix \ref{sec:AppendixNorms}, and are fixed to satisfy
\be
\frac{1}{N_{nlJ}^\CO} 
=\sqrt{ \frac{{\rm vol} \left( S^{d-1} \right) \Gamma(\Delta + n +l) \Gamma \left( \Delta + n - \frac{d-2}{2} \right)
\Gamma \left(\frac{d}{2} \right)}{ n! \Gamma(\Delta) \Gamma \left(\Delta - \frac{d-2}{2} \right)
\Gamma \left(\frac{d}{2} + n + l  \right)}}
\label{eq:Onorm}
\ee
Note that the existence of $\CO$ in the dual CFT is easy to prove 
constructively in this approach: one can directly obtain $\CO(t, \hat{x})$ by taking the limit of $\phi(t, \hat{x}, \rho) \cos^{-\Delta} \rho$ as $\rho \to \pi/2$.  
We have quantized $\phi$ and $\CO$ using identical creation and annihilation operators, so we can directly translate CFT states constructed by functions of $\CO(x)$ into bulk AdS states, and vice versa\footnote{This equivalence may also be useful in clarifying the behavior of precursor states in AdS/CFT \cite{Giddings:2001pt,Freivogel:2002ex,Hubeny:2002dg,Hubeny:2000eu}.}.  In particular, we can understand how to use $\CO(x)$ to create bulk AdS states that look like flat space in and out states near the center of AdS.  

With our explicit construction, it is easy to create one-particle states localized in AdS by using the operator $\phi(x)$.  For example, if we want our particle to have a bulk wavefunction $\psi(\hat{x}, \rho)$ at time $t$, we simply integrate $\phi(x) | 0 \>$ against an appropriate projection function.
However, via the AdS/CFT correspondence, we can also obtain this state using only the CFT operator $\CO(x)$, but we must smear $\CO(x)$ over the entire boundary spacetime -- we cannot isolate states if we only act with $\CO(x)$ at a single time.  The most elementary example involves isolating the state created by $a_{nlJ}^\dag$, which we can do using the trivial boundary smearing 
\be
| nlJ \rangle \propto \int dt d^{d-1} \hat{x} \ \!  e^{i E_{nl} t} Y_{lJ}(\hat{x}) \CO(t, \hat{x}) |0 \rangle .
\ee
Yet without the time integral, we would necessarily include all values of $n$.
So to select in and out states in the flat space limit, we will need to smear CFT operators over time intervals that are very large compared to their inverse energies.  

\begin{figure}[t!]
\begin{center}
\includegraphics[width=0.65\textwidth]{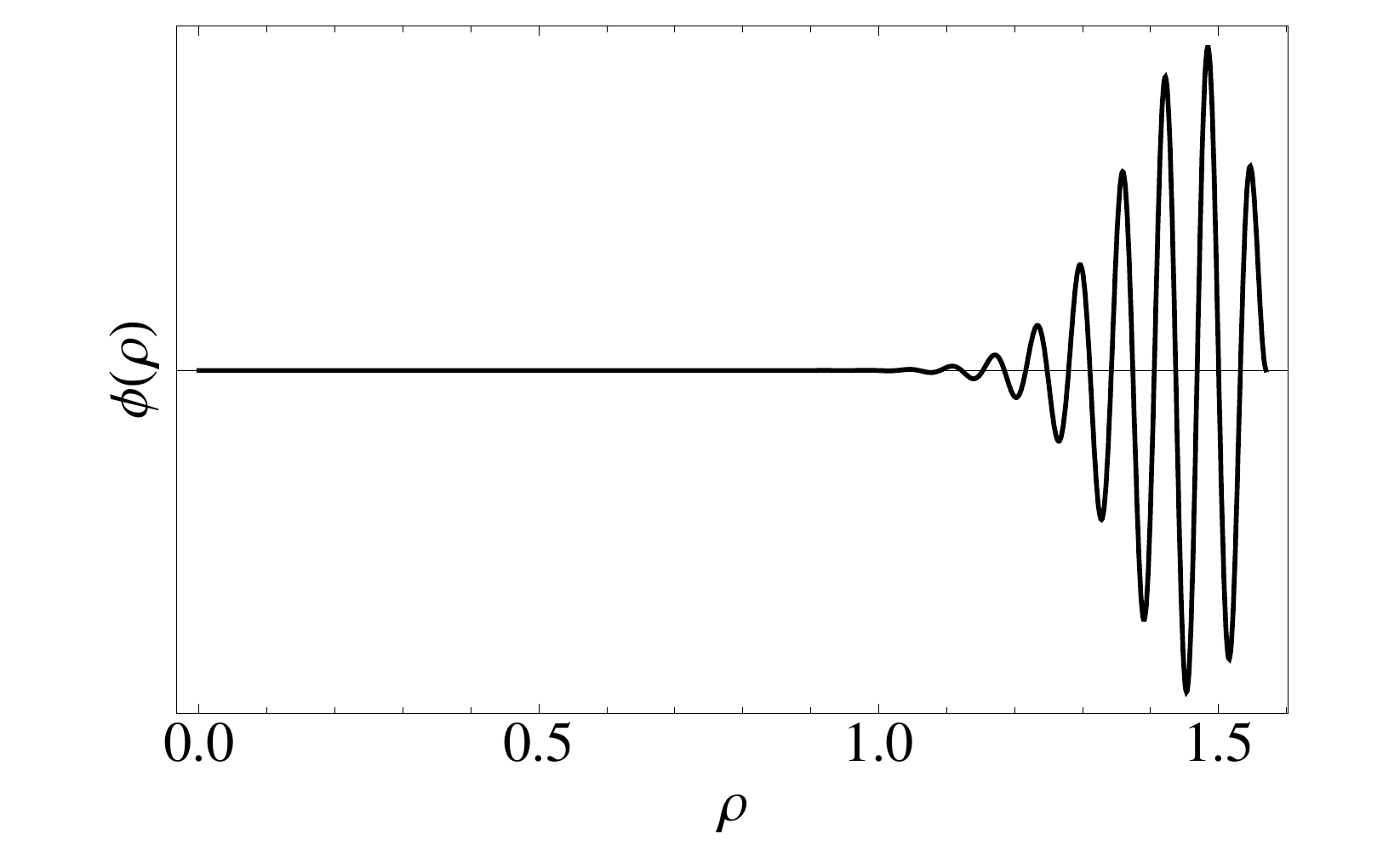}
\caption{ The wavefunction $\< 0| \phi(t,x) | \psi_{\omega, \tau} \rangle$ at $t=0$ corresponding to the state $| \psi_{\omega, \tau} \rangle$ constructed in eq. (\ref{eq:smeared}) by smearing the single-trace operator $\CO(t,\hat{x})$ acting on the vacuum.  The choice of energy is $\omega = 100$ and of smearing window size is $\tau = 0.1$, in units of the AdS radius.  The state is localized near the boundary, though the wavefunction falls to zero at $\rho=\pi/2$, as is necessary for a normalizable mode.   
\label{fig:smeared}}
\end{center}
\end{figure}

Now we will consider a couple of more interesting examples. 
 If we smear $O(t, \hat x)$ in a region of size $\tau$ where $\frac{1}{\tau} \gg \Delta$ with a phase factor to pick states with large energy $\omega \gg \frac{1}{\tau}$, then we obtain
\be
| \psi_{\omega, \tau} \rangle = \left[ \int_{-\infty}^\infty dt e^{-i \omega t - t^2/\tau^2} \int d^{d-1} \hat{x}
\CO(t, \hat{x}) \right] |0 \rangle .
\label{eq:smeared}
\ee
This integral will be dominated by terms in the mode expansion of $\CO$ with $2n \approx \omega$, and with a width about this $n$ of size $1/\tau$.  
If we examine the wavefunction $\< 0 | \phi(t,x) | \psi_{\omega, \tau} \>$, we find that it describes a particle created a distance of order $\tau$ from the boundary at $\hat{x}_0$,
as depicted in Figure \ref{fig:smeared}. This particle will subsequently fall into the bulk,  and then oscillate back and forth through AdS, making its closest approach at  a distance $\tau$ from the boundary.  This picture is very familiar from other studies of AdS/CFT.

A less trivial example is provided by a particle of large mass created in a non-relativistic state near the center of AdS.  For simplicity, we can prepare the particle in an s-wave
\be
|\psi_{NR} \> = \int_{-\tau}^\tau dt d^{d-1} \hat{x}  e^{i \left( \Delta + K \right) t} \CO(t, \hat{x}) |0 \> ,
\ee
where we take the limit that $\Delta \gg K \gg \frac{1}{\tau} \gg 1$.  In this limit, there seems to be a paradox with causality when we measure the wavefunction $\< 0 | \phi(t,x) | \psi_{NR} \>$.  Our particle has a large mass and small momentum, so by energy conservation it cannot exist anywhere except very close to the center of AdS.  But since the operator $\CO$ only acts for a very short time, the particle appears long before a signal can propagate from the boundary to the center of AdS.  

In fact there is no problem with causality, since we cannot use this mechanism to send a signal faster than light.   This phenomenon is possible because the two point function $\< 0 | \phi(x) \CO(y) | 0 \>$ does not vanish outside the lightcone, but this behavior should be familiar from flat space field theory.  In fact, the related two point function $\< 0 | \phi(x) \phi(y) | 0 \>$ does fall off as $e^{-\Delta \sigma(x,y)}$ outside the lightcone, where $\sigma(x,y)$ is the geodesic distance between $x$ and $y$.  Since $\CO(y)$ is proportional to $\phi(y,\rho)$ rescaled by the inverse of this exponential factor and taken to the boundary at $\rho \to \pi/2$, it should be expected that $\CO$ can act on the vacuum and  create a particle that is instantly deep in the bulk of AdS. 

The standard bulk-boundary propagator not causal, and it can be directly computed in our language as
\be
G_{\partial B}(t,\hat{x}, \rho; t', \hat{x}') = \< 0 | \phi(t,\hat{x},\rho) \CO(t', \hat{x}')  |0 \> \propto 
\left( \frac{e^{i t'} \cos \rho}{\cos(t-t') - \sin \rho \hat{x}\cdot \hat{x}' }
\right)^\Delta .
\ee
One can use this function to compute the $\phi(x)$ wavefunction created by a source function $f(y)$ on the boundary by integrating $G_{\partial B}(x, \rho; y)$ against $f(y)$ \cite{Hamilton:2005ju}.  As is familiar from flat spacetime, there are advanced, retarded, and Feynman versions of $G_{\partial B}$ that correspond to the different possible time orderings of $\phi(x,\rho)$ and $\CO(y)$.  

Let us now consider which bulk states are localized in our universe at the center of AdS, and then we will determine how to create and manipulate them using CFT operators.   AdS contains an infinite number of regions of size $R$, but in the flat space limit we take $R \to \infty$, and a single one of these regions will become our flat-space universe.  As a maximally symmetric space, AdS does not have a preferred center, but for our purposes we can define $\rho = 0$ as the origin and focus on a `universe' of size $R$ around it.  We can then ask which states are localized in this universe.   Among one-particle states, the $n,l=0$ state created by $a_{000}^\dag$ has the largest support at small $\rho$.  As one can see by inspection of equation (\ref{eqn:phinlj}), this state has a wavefunction proportional to $\cos^\Delta \rho$, so that it decreases rapidly for large $\rho$, especially as $\rho$ approaches the boundary at $\pi/2$.  In contrast, states with larger $n,l$ extend further away from the origin of our universe.  We will now investigate the multi-particle equivalents of this $n=0$ state; we will find that they are naturally equated with the flat-space in and out states of zero momentum.

\subsection{Large Energy Primaries as In and Out States}
\label{sec:primaries}

In the previous section we distinguished the one-particle $n,l=0$ state due to its localization in our universe at the center of AdS.  This state also plays a very special role in the CFT -- it is primary.  To review what this means, consider the conformal algebra
\be\label{eq:confalgebra}
\left[ M_{\mu\nu} , P_\rho \right] = i (\eta_{\mu\rho} P_\nu - \eta_{\nu\rho} P_\mu)
, && \left[ M_{\mu\nu} , K_\rho \right] = i (\eta_{\mu\rho} K_\nu  - \eta_{\nu\rho}
K_\mu) , \nn\\
\left[ M_{\mu\nu} , D \right] = 0 , && \left[ P_\mu, K_\nu \right] =
- 2 (\eta_{\mu\nu} D + i M_{\mu\nu} ) , \nn\\
\left[ D, P_\mu \right] =  P_\mu , && \left[ D, K_\mu \right] = - K_\mu .
\ee
The dilatation operator $D$ and a subset of the rotation generators $M_{\mu \nu}$ form the largest commuting subalgebra; we used this fact above to quantize $\phi$ and $\CO$ in terms of the $a_{n l J}$, which annihilate states of definite dimension and angular momentum.  For the remainder of this section we will work with states of definite dimension and angular momentum, which will correspond to spherical waves in the flat space limit.

On this space of states, $K_\mu$ and $P_\mu$ act as lowering and raising operators, respectively, and the primary states are those annihilated by $K_\mu$.  From the commutation relations we see that a primary state will remain primary when acted on by $D$.  So in other words, since time translations are generated by $D$, the property of being primary is conserved.  We will have more to say about the conformal algebra soon, but for the moment let us construct multi-particles states that scatter in our universe.

We saw above that among one-particle states, the primaries are the ones most localized in our universe.  This feature of primary states generalizes to multi-particle states, so that all primaries have their center of mass at the center of AdS.  To see this, consider a two-particle primary state $| \Psi \rangle$.  The bulk wavefunction for this state is
\be
\psi(x,y) = \langle 0 | \phi(x) \phi(y) | \Psi \rangle ,
\ee
and in general the two particles will move around in AdS, perhaps passing in and out of our universe.  We can determine whether these two particles pass through our universe simultaneously by computing the wavefunction of their center of mass, $\psi(x,x)$.  Since $|\Psi \rangle$ is primary, $K_\mu \psi(x,x) = 0$, and using the explicit form of $K_\mu$ in AdS allows us to derive that
\be
\psi(x, x) \propto \left( e^{it} \cos \rho \right)^E ,
\ee
where $E$ is the eigenvalue of $|\Psi \rangle$ under the dilatation operator.  We see that for large $E$, the center of mass of our state localizes very sharply near the center of AdS, at $\rho = 0$, and falls very rapidly near the boundary of AdS, where $\rho = \pi/2$.  This result generalizes to primaries with many particles.

We can use this information to explicitly construct a set of in and out states that scatter in our universe using only CFT operators in the large $N$ limit.  If we begin with a primary single-trace operator $\CO(x)$, then we can construct an $m$-particle state with the operator
\be
\mathcal{C}_f = \int \prod_{i=1}^n d^d x_i f(x_1,x_2,...,x_m) \CO(x_1)... \CO(x_m) .
\ee
One might object that this integral develops singularities when the points $x_i$ approach each other, corresponding to processes where particles interact near the boundary of AdS.  In cases where one can quantize $\CO(x)$ directly, these should be eliminated by normal ordering the $\CO(x_i)$.  In general one should subtract from the product of the operators all of the singular terms in the various OPEs, leaving a manifestly finite result.  
The singular terms correspond to AdS processes where the particles in the initial or final states interact while they are close to the boundary, and as such they would be related to disconnected scattering processes.

Imposing the condition that ${\cal C}_f$ is a primary operator provides an equation for the function $f$.  This equation will depend only on the action of the conformal generators, and thus it should be valid non-perturbatively as long as we use the exact $\CO(x)$.  Let us see how this works in the simple case where $d=2$, $m=2$.  The operator $\mathcal{C}_f$ will have dimension $\omega$, and angular momentum $L$.  This means we can write
\be
\mathcal{C}_f = \int d^2x_1 d^2x_2 \CO(x_1) \CO(x_2) e^{i \omega (t_1+t_2)/2 + i L (\varphi_1 + \varphi_2)/2} f(t_{12}, \varphi_{12}) ,
\ee
where $t_{12} = t_1 - t_2$ and $\varphi_{12} = \varphi_1 - \varphi_2$.   Now the condition that $\mathcal{C}_f$ is primary requires that
\be
0  = \int d^2 x_1 d^2 x_2 \left( [K_\mu, \CO(x_1)] \CO(x_2)  + \CO(x_1) [K_\mu, \CO(x_2)]  \right) e^{i \omega (t_1+t_2)/2 +  i L (\varphi_1 + \varphi_2)/2} f(t_{12}, \varphi_{12}) . \nonumber 
\ee
The commutators of the special conformal generator with $\CO(x_i)$ reduce to differential operators acting on $\CO(x_i)$, and these can be integrated by parts to give an equation for $f$.  In two dimensions, there are two special conformal generators, which act as $K_\pm = i e^{-i t \pm i \varphi} ( \d_t \mp \d_\varphi )$.  After some simplification, we obtain the equations
\be
\frac{ \d_{t_{12}} f \pm \d_{\varphi_{12}} f }{f} &=& \left( \frac{\omega \pm L}{2} 
\right) \cot \left( \frac{ t_{12} \pm \varphi_{12} }{2} \right).
\ee
These two equations determine that
\be
f(t_1, t_2, \varphi_1, \varphi_2) =  e^{i \omega(t_1 + t_2)/2 + i L (\varphi_1 + \varphi_2) /2} \sin \left( \frac{t_{12} + \varphi_{12}}{2} \right)^{(\omega + L)/2}
 \sin \left( \frac{t_{12} - \varphi_{12}}{2} \right)^{(\omega - L)/2}.
\ee
up to an overall normalization factor.
We can use the binomial theorem to expand these powers of sine into a sum of exponentials, and it turns out that this mimics the expansion of the standard radially quantized double trace operator $\CO \partial_\mu^L \partial^{2 n} \CO$ in derivatives, term-by-term, satisfying our expectations for what a double trace primary operator should be.  One can straightforwardly extend this procedure to $m$-particles states and $m$-trace operators in $d$ dimensions.  It is worth noting that when we include more particles in higher dimensions, the $f$ functions are not completely determined by the primary condition.  The number of remaining degrees of freedom in $f$ precisely coincides with what we would expect for a state of $m$ particles in $d$ dimensions constrained to be on-shell with zero total spatial momentum.  

In the flat space limit we take $R \to \infty$, which means that the bulk energy, or CFT dimension $\omega$, of our scattering states will become very large in units of $1/R$.  In the limit that $\omega \to \infty$ with fixed $L$, our expression for $f$ simplifies to
\be
f(t_1, t_2, \varphi_1, \varphi_2) \approx  e^{i \omega(t_1 + t_2)/2 + i L (\varphi_1 + \varphi_2) /2} e^{-\omega \left( t_{12}^2 + (\varphi_{12}-\pi)^2 \right)/8}.
\ee
Thus $f$ has vanishing support unless the operators $\CO(x_1)$ and $\CO(x_2)$ act at identical times and at antipodal points on the boundary of AdS.  This is precisely what we would expect for a high energy two-particle state with zero total spatial momentum.  In addition, the uncertainty in the energy of these particles will be of order $\sqrt{\omega}$, so the relative uncertainty in the energy, $1/\sqrt{\omega}$, vanishes in the flat space limit.

We have found the exact primary states of the CFT, which involve particles passing back and forth through our universe at the center of AdS, but these states exist for all time.  They cannot be used as in and out states because they would experience an infinite number of scattering processes, as we will see explicitly in section \ref{sec:multiplescat}.  To avoid this problem we need to regulate the time integrals defining $\mathcal{C}_f$, preparing the in states at a definite time and collecting the out states after the scattering process has occurred.  This means we will introduce a parameter $\Delta T$ and write
\be
\label{eqn:GeneralState}
\mathcal{C}_f = \int \prod_{i=1}^n \left[ d^d x_i R\left(\frac{t_i}{\Delta T} \right) \right] f(x_1,x_2,...,x_m) \CO(x_1)... \CO(x_m) ,
\ee
where $R(a)$ is a regulator function such as a Gaussian.  We should take $\Delta T > 1/\sqrt{\omega}$ so that the regulator does not decrease the energy resolution of our states, and we need $\Delta T \ll R$ so that the in and out states are near the boundary and are therefore not interacting.  

The regulator that we have imposed on the integration times is essential to our definition of the in and out states.  To have a non-trivial scattering matrix, the in and out states must represent different bases for the Hilbert space of scattering states; this occurs in our setup because the in and out states will be centered about different times.  Furthermore, the in and out states should each diagonalize the hamiltonian (dilatation operator), and this is only possible because $\Delta T \ll R$, so that the particles in these states are very far apart.

It is worth emphasizing that the technical complications that we have discussed above are almost identical to those that one encounters while defining the S-Matrix directly in flat spacetime.  Subtracting out the disconnected components in the OPE is equivalent to only considering the connected components of the flat space S-Matrix.  Corrections from interactions are a problem in AdS/CFT because they obscure the distinction between single and multi-particle states in the guise of single and multi-trace operators.  In flat space this issue gets resolved for stable massive particles by isolating them at rest and then boosting them to infinity, while for massless particles this problem is not resolved -- amplitudes with definite numbers of massless particles generically lead to IR divergences, and force us to ask inclusive questions about objects such as jets.  In AdS the situation is actually much simpler -- we can isolate massless particles, since they have a minimum energy of $\CO(1/R)$.  However, in the flat space limit the IR divergences return, because $1/R$ was serving as an IR regulator.

Before moving on to discuss the flat space limit of the conformal algebra, let us pause to consider the generality of these results.  
In any CFT one can perform the construction of equation (\ref{eqn:GeneralState}) and obtain $\Delta T$-regulated primary operators from individual primaries $\CO_i$ that have definite dimension under the full Dilatation operator $D$.  However, the composite operator $\mathcal{C}_f$ constructed in this way will not have definite dimension.  When evolved by $D$ it will mix with all of the other operators in the CFT, and so we will be able to define a mixing matrix by taking inner products of distinct regulated primary states created by $\mathcal{C}_{f_1}$ and $\mathcal{C}_{f_2}$ that are centered at different times.  This mixing matrix is not yet an S-Matrix because in a general CFT we have no way of selecting the set of preferred operators $\CO_i$ with which to begin forming the in and out states created by $\mathcal{C}_f$.  At large $N$ there is a natural choice: the single trace primary operators.  Note that the `t Hooft coupling $\lambda$ has not made any direct appearance in our construction; the large $\lambda$ limit will be necessary if we want a theory with a small number of light particles, but it may be interesting to examine our S-Matrix construction at small $\lambda$ in future work.

\subsection{Flat Space Limit of the Conformal Algebra and Plane Waves}
\label{sec:planewaves}

Now that we have constructed CFT states that scatter in our universe, let us consider what happens to the conformal algebra in the flat space limit.  This limit comes about by defining the Hamiltonian, momentum, and boost generators as
\be
H_f &=& \frac{1}{R} D , \\
P_f^i &=& \frac{i}{2R} (P^i-K^i) , \\
B_f^i &=& \frac{1}{2}(P^i + K^i),
\ee
where $R \to \infty$ in flat space, and we have replaced $\mu \to i$ to emphasize that $P_f^i$ and $B_f^i$ are the spatial momenta and boosts.  To leading order in $1/R$, we find the commutation relations
\be
[H_f, B_f^i] = -i P_f^i, \ \  [P_f^j, B_f^i] = -i \delta^{ij} H_f, \ \  [H_f, P_f^i] = 0 ,
\ee
as expected for the hamiltonian, the spatial momenta, and the boost generators in flat space.  The CFT operator $M_{ij}$ turns into the generator of rotations, and requires no rescaling.  The first two commutation relations are exact, while in the last we have neglected $B_f/R^2$.  In the $R \to \infty$ limit, the discrete spectrum of the raising and lowering operators $P$ and $K$ becomes a continuum, in the same way that the discrete spectrum of states in a box becomes continuous as the box becomes infinitely large.

For any finite value of $R$, the flat space limit eventually breaks down.  For example, we can translate a state out of our universe by applying $e^{i H_f T}$ or $e^{i P_f^i X_i}$ with $T$ or $X_i$ of order $R$, at which point the flat space commutation relations become invalidated.  However, since the commutations
with boosts are exact, we do not run into any problems with very large boosts, or with states with large energy and momentum.  This accords with our intuition, since the breakdown of flat space should only be visible when we traverse distances of order the curvature scale.

We have seen that the primary states of a CFT play a special role in the flat space limit $R \to \infty$.  In that limit $K = iR P_f + B_f$, and so primaries have vanishing total flat-space momentum.  Since in the flat space limit momentum is conserved, all non-trivial S-Matrix elements can be obtained from matrix elements of the primary states.  However strictly speaking, we could also discuss the scattering of states that are almost primary, meaning that they have been translated from the origin of our universe by a distance small compared to $R$, or that their total momentum is small but non-zero.

Thus far we have seen that regulated primary multi-trace states with large dimension are a very natural and sharply-defined sector in the CFT that matches onto spherical waves in the flat space limit. 
However, we are most familiar with amplitudes computed in terms of single-particle plane waves, and it would be convenient to be able to work directly with such states in the CFT.  
AdS does not enjoy a symmetry under translations, so plane waves are necessarily deformed by the AdS curvature.   However, we will give a precise CFT construction of states that behave like plane waves near the center of AdS, which is all we need to compute the S-Matrix in the standard way.  

We can construct states that look like plane waves near the center of AdS by using our explicit knowledge of the parallel quantization of bulk and boundary operators.  We will proceed by using equation (\ref{eqn:phiflat}), which shows that in our universe at the center of AdS, the bulk wavefunctions reduce to flat space wavefunctions.  Since we know that in spherical coordinates
\be
e^{i \vec p \cdot \vec x} \propto \sum_{lJ} j_l(px) Y_{lJ}^*(\hat{p}) Y_{lJ}(\hat{x}) ,
\ee
we can find a generalization of $e^{i p \cdot x}$ to all of AdS by simply replacing the spherical Bessel function with an appropriate hypergeometric function.   This immediately tells us what linear combination of bulk wavefunctions we need to make a plane wave, and then we can translate this information to give an explicit CFT operator that constructs these states.  To see how this works, consider a plane wave state defined by $| p \rangle = \sum c_{n l J} a_{nlJ}^\dag |0 \rangle$ so that
\be
\langle 0 | \phi(x) | p \rangle = \langle 0 | \phi(x) \left( \sum_{nlJ} c_{n l J} a_{nlJ}^\dag |0 \rangle \right)  \approx e^{i p \cdot x} 
\ee
for $\rho \ll 1$.  We can easily determine $c_{nlJ}$; first note that to have a fixed energy $E = \Delta + 2n + l$ we must relate $n$ and $l$, so we really need only sum over $l$, and we will henceforth label states with $E$ and $l$ quantum numbers. So to obtain a plane wave in the flat space limit, we just take $c_{E l J} = Y_{lJ}^*(\hat p)$.   Now we want to find a function $C_p(t, \hat x)$ so that
\be
\label{eqn:pstatefromO}
\int dt d^{d-1} \hat x \ \! C_p(t, \hat x) \CO(t, \hat x) | 0 \rangle = |p\rangle .
\ee
Using our explicit quantization from section \ref{sec:hilbertspaces}, we see that 
\be
C_p(\hat x, t) =  e^{i E t} \sum_{l J}  \frac{1}{N^\CO_{\Delta E l}} Y_{lJ}^*(\hat p) Y_{lJ}(\hat x) .
\ee
This appears to be a very complicated function, but luckily in the limit that $E \gg l$, $N^{\CO}_{E l}$ is actually independent of $l$, so that
\be
C_p(\hat x, t) \propto  p^{m-\frac{d}{2}} \left( \frac{E+m}{E-m} \right)^{\frac{E}{2}} \frac{m^{\frac{d-4m}{4}}}{2^m}  e^{i E t} \delta^{d-1}(\hat p, \hat x) 
\propto \left\{ \begin{array}{ll} \left( \frac{m}{p^2} \right)^{d/2} & m \gg p \\
\frac{1}{E^{d/2}} & m=0  \end{array} \right\}. .
\ee
where the sum over the spherical harmonics has been reduced to a spherical delta function, setting $\hat p = \hat x$.  This result accords with those of \cite{Gary:2009ae}.

Now let us derive this result in a different way that will allow us to see its limitations, using the correspondence between the conformal and Poincar\'e algebras.
Having identified the Poincar\'e generators in the flat space
limit of AdS,
we can attempt a precise construction of plane wave single-particle
states, in terms of the exact energy- and spin-eigenstates of global AdS.
We will use AdS$_3$ to build intuition in a simple setting, 
since it will allow us to use
some simple explicit expressions for the action of the generators.
 
For the AdS$_3$ case, where we can write the states in terms of their holomorphic and anti-holomorphic quantum numbers, so that $m-\bar{m} =l$ and $m+\bar{m} = 2n + |l| = E-\Delta$.  We can write the flat space momentum operator as $P^\pm_f = P^1_f \pm i P^2_f$ and use the fact that  $P^+, K^-$ act only on $m$ and $P^-,K^+$ act only on $\bar{m}$. In the limit of large $E$ we find
\be
K_{\pm} | E, l \> &\approx& |p| |E+1, l \pm 1 \>, \ \ \ \ \ P_{\pm} | E, l \> \approx |p| |E - 1, l \pm1 \> ,
\ee
where $|p| = \sqrt{E^2 - \Delta^2}$.  For large $E$,  the states of energy $E$ and $E + 1$ are very similar, so we guess that our plane wave momentum state will be
\be
| p \> &=& \frac{1}{N} \sum_{\omega,l} (-1)^{-\frac{\omega}{2}} Y_l(\hat{p}) f_E(\omega) |\omega, l\>,
\ee
where $f_E(\omega)$ is an energy spread function that is sharply peaked at $\omega = E$, but broad on scales of order $1$ (all energies are measured in units of $1/R$).
A quick calculation shows that
\be
P_f^\pm | p \>&=&
   i |p| \frac{1}{N} \sum_{\omega,l}  (-1)^{-\frac{\omega} {2}} Y_l(\hat{p})f_E(\omega) {1 \over 2} \left( |\omega-1, l\pm 1\> -   | \omega+1, l \pm 1\> \right)  \nn\\
  &\approx& p^\pm | p \> ,
\ee
as desired for a momentum eigenstate.  We can build these functions as in equation \ref{eqn:pstatefromO} by using precisely the function $C_p(t, \hat x)$ above, except that in order to achieve an energy smearing such as $f_E(\omega)$, we must multiply $C_p(t, \hat x)$ by a suitable time smearing function with slow time variation, such as the fourier transform $\tilde f_E(t)$. 

\subsection{Local Particles and the CFT}

One of the mysterious aspects of the AdS/CFT duality is that locality emerges in the holographic direction. 
Naively, one would not guess that it is possible to create `in' and `out' states consisting of well separated particles, because the boundary theory lives on a circle of fixed finite radius (or perhaps more accurately, the boundary theory is conformal, so distances do not mean anything).  
However, we know from the bulk description that at large $R$ the center of AdS looks like flat-space,
and we can study particles that are very far apart from each other and therefore very weakly interacting. 
As we have emphasized, any state in AdS is also a state in the CFT, so
there must exist CFT states with sharply localized AdS dual positions.
Such states can clearly be far apart from each other in a meaningful
sense, so that their radial evolution is approximately that of
an infinite $N$ CFT.\footnote{Of course, there is still the issue of
whether a general large $N$ CFT has an AdS dual whose interactions are
local, which we do not address in this paper.}  
What is somewhat mysterious though is what such states look like from the
CFT point of view.

\begin{figure}[t!]
\begin{center}
\includegraphics[width=0.85\textwidth]{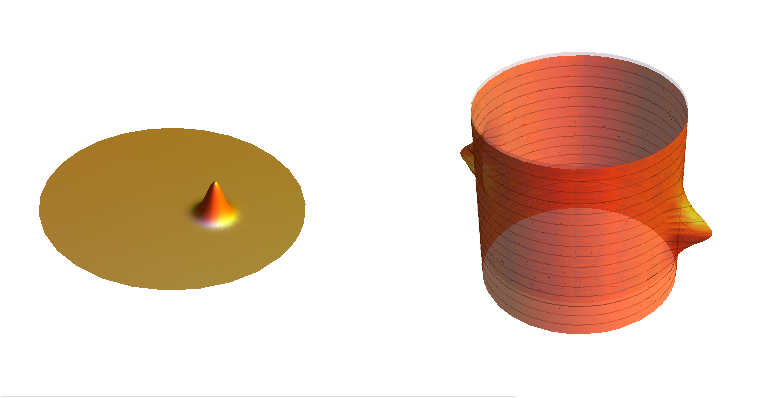}
\caption{Left: The bulk wavefunction $\< 0 | \phi(x) | \psi\>$ at $t=0$, of a 
non-relativistic particle state $|\psi\>$ with
a gaussian wavepacket, near the center of AdS$_3$. Right: The corresponding
wavefunction $\<0| \CO(x) | \psi \>$ in the boundary theory, plotted along the
surface of the boundary cylinder. Time runs
upward along the cylinder, and the magnitude $|\< 0 |\CO(x) | \psi\>|$
of the wavefunction is its extent outward from the cylinder surface. 
Knowledge of the bulk wavefunction $\phi(x)$ and $\dot{\phi}(x)$ everywhere
in AdS at a given time is enough to determine the state; by contrast,
one needs the boundary wavefunction $\CO(x)$ at all times in order to
extract the same information.
\label{fig:bulkboundwvfn}}
\end{center}
\end{figure}

In this section we will briefly examine the extreme example of a heavy non-relativistic particle state $|\psi\>$ with large
mass $m$ and small momentum $\vec{k}$, whose wavefunction is well-localized to the position $x(t)$ near the center of AdS. 
We will take the bulk wavefunction to be a gaussian with spatial and momentum
spread $\Delta x$ and $\Delta k$ respectively.  We make the approximations
\be
m \gg k \gg \frac{1}{\Delta x} \gg \frac{1}{R} \ \ \ \mathrm{and} \ \ \ (\Delta x)^2 m \gg R .
\ee
The first gives us a well-localized, non-relativistic wavefunction in the flat-space limit, while the latter means that the wavefunction spreading is negligible even after times of order $R$.  The curvature of AdS has no effect on the particle until it reaches distances of order its small velocity $k/m$, at which time it will be deflected back in toward smaller radii. We will restrict our attention to time-scales shorter than this so that all curvature effects on the particle trajectory are negligible. We can describe our state as
\be
|\psi\> \approx \int d^{d} p \psi(\vec{p}) |\vec{p}\ \> , \ \ \ \ \ \ \ 
\psi(\vec{p}) &\propto& \exp\left( - \frac{(\vec{p}-\vec{k} )^2}{\Delta k^2} + 
 i x_0 \cdot \vec{p} \right),
\ee
 Its bulk wavefunction is then simply
\be
\psi_\phi(x) = \< 0 | \phi(x) | \psi \> \propto
\exp \left( - \frac{(\vec{x} - \vec{v} t -x_0)^2 }{\Delta x^2 } - i \vec{k} \cdot
\vec{x} + i \omega_k t\right),
\ee
where $\vec{v} \equiv \vec{k}/m$ and $\omega_k \approx m + k^2/2m$.  
Now we can measure the boundary wavefunction $\psi_\CO(x) \equiv
\< 0 | \CO(x) | \psi \>$.  From equation (\ref{eqn:pstatefromO}) we have 
 \be
 \< 0 | \CO(x) | \vec{p} \> \propto \frac{\delta(\hat{p},\hat{x}) e^{-i \omega t }} {N^{\CO}_{\omega,m}}
 \ee 
 In the non-relativistic limit $p \ll m$ in $\omega_p$, we find the leading  $p,m$-dependence is simply
\be
\frac{1}{N^{\CO}_{\omega, m}} &\propto& \frac{m^{d/4}}{p^{d/2}}
\ee
and does not depend on $l$ except through the combination $p^2 = 2(2n+l)m/R$.  We can now derive the boundary wavefunction for this state, it is
\be
\psi_\CO(x) &=& \< 0 | \CO(x) |\psi \> \propto
 \int d^d p  \left( e^{-\frac{(\vec{p}-\vec{k} )^2}{\Delta k^2} +  i \vec{x}_0\cdot \vec{p} } \right)  \left( \frac{\delta(\hat{p}, \hat{x} )}{p^{d/2}} e^{-i (m+ \frac{p^2 }{2m})t} \right) \\
 &\propto& \exp\left[- i mt  - \frac{k^2 \sin^2 \theta}{\Delta k^2} - \frac{ \left( \hat{x} \cdot ( \vec{x}_0 + \frac{\vec{k}}{m} t  ) \right)^2 }{\Delta x^2} +i \hat{x}\cdot \left(\vec{x}_0 + \frac{\vec{k}}{2m }t \right)  \cos \theta \right]  
 \nn
\ee
where we have defined $\hat{k} \cdot \hat{x} \equiv \cos \theta$.   Note that this wavefunction is exponentially suppressed except when the direction $\hat{x}$ of the position on the boundary aligns with the momentum of the particle.  It is also exponentially suppressed when the  position $\vec{x}_0 + \vec{k} t/m$ in the bulk has a large component along $\hat{x}$.  In order for both of these suppressions to be avoided, then, one must be looking at the boundary wavefunction at angles close to $\hat{k}$ and at the time when the particle makes its closest approach to the center of AdS.  This is depicted in Figure \ref{fig:bulkboundwvfn}.

Locality and AdS/CFT have been studied in \cite{holographyCFT,Heemskerk:2010ty,Heemskerk:2010hk,Giddings:2000ay,ArkaniHamed:2000ds}, but it might be interesting to further explore the consequences of locality in this non-relativistic limit, where one could consider the scattering of two heavy charged particles as they exchange a massless photon.  The formation of bound states might be particularly interesting to examine as a CFT process.

\sect{Scattering Amplitudes from a CFT}

\subsection{Parametrics of the Flat Space Limit and the S-Matrix}
\label{sec:Parametrics}

Although many precise examples of the AdS/CFT duality are known, one can also investigate the general properties of low-energy effective field theories in AdS space, and interpret the results in terms of a putative CFT.   In this way, we can study the AdS/CFT correspondence in a very general and robust way, because EFTs almost always provide a good description of both field theories and string theories at low-energy.  We will couch our discussion of the S-Matrix in the language of AdS effective field theory.

Let us consider the scales in the AdS effective theory.  In the flat space limit, the curvature of AdS must become negligible, so the AdS length $R$ will be our largest distance scale.  Even without gravity, a theory in AdS can be used to construct representations of the conformal algebra, but we must include gravity to expect the boundary theory to be a CFT with a local energy-momentum tensor.  So the shortest distance scale available will be the Planck scale, $\ell_{pl}$, which is parametrically separated from $R$ by a power of the parameter $N$ in the large $N$ CFT dual.  If there is a string scale $\ell_{s}$ then it too must be parametrically separated from $R$, by taking the limit of very large 't Hooft coupling $\lambda$.  Thus in the canonical example of AdS$_5$, we have the three energy scales $1/R \ll \lambda^{1/4}/R < N^{2/3}/R$.  Compactification radii might be kept small and therefore remain invisible to the effective field theory, or they could be of order $R$, decompactifying in the flat space limit.  Extra compact dimensions of size $R$ are very well motivated in supersymmetric examples of AdS/CFT, but for our general and not necessarily supersymmetric setup we will ignore them, keeping only the AdS space.  Thus to take the flat space limit, we require $N \to \infty$, and if a string scale is present, we must take $\lambda \to \infty$ as well, as discussed long ago in \cite{Polchinski:1999ry,Susskind:1998vk}. 

Next we will consider what happens to the quantum numbers of interesting states in the flat space limit.  The energies of states in AdS are given as multiples of the AdS scale $1/R$, so we must send $ER \to \infty$ to obtain sensible flat-space states.  Similarly, if we wish to describe massive particles, we must send $\Delta \approx m R \to \infty$ so that $\Delta/R$ remains fixed.   However, since angular momenta are dimensionless, we can keep $l \ll n$ finite.
Note that if all of the couplings vanish as $1/N$, then we also require the limit $ER \to \infty$ in order to have a non-trivial S-Matrix.

We cannot explore physics at distances near the Planck length in effective field theory, but we can still use EFT to setup incoming and outgoing states for a trans-planckian scattering processes.  This means that there should be no obstruction to discussing well-separated single-particle states boosted to energies larger than the Planck scale.  We can use our methods to setup CFT states that will develop into `small', sub-AdS scale black holes, which we expect to decay back into Hawking radiation that is well described by the bulk EFT.  If CFTs at large $N$ and large $\lambda$ could be solved exactly, we could use the solution to understand the mechanism by which information is transferred to Hawking radiation and unitarity preserved.

\begin{figure}[t!]
\begin{center}
\includegraphics[width=0.5\textwidth]{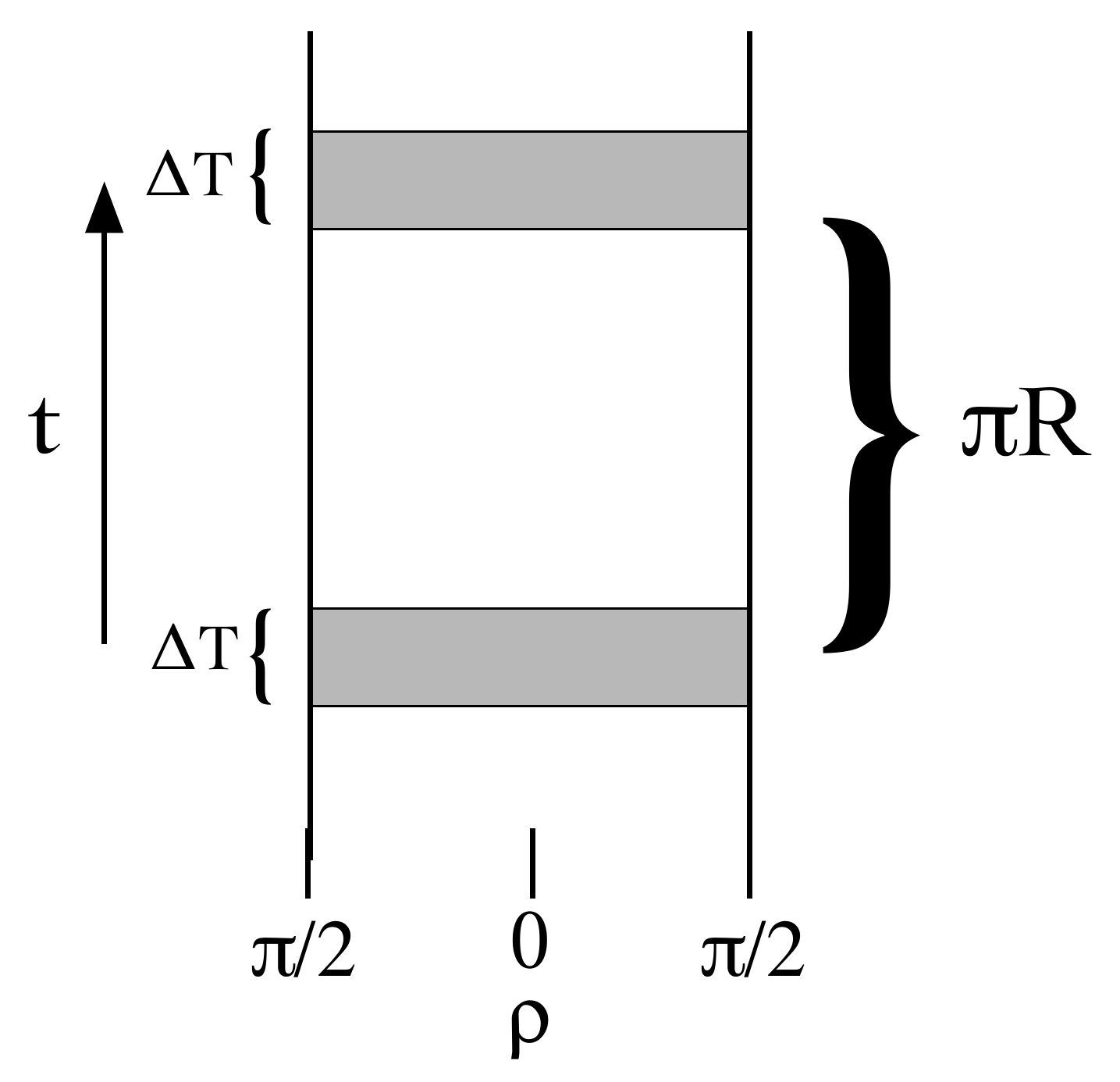}
\caption{  In this figure we have a flattened version of the AdS cylinder.  The in and
out states are prepared by integrating CFT operators over the grey regions with the smearing functions discussed in section \ref{sec:States}. 
Particles in AdS take a time $\pi R$ to travel from their closest approach to the boundary through the center of AdS and back out to the boundary, so we separate the in and out states by this time interval. We take $\Delta T \ll R$ so that the particles are created near the boundary, where they do not interact, and $\Delta T$ much larger than their wavelengths, so that their energies and momenta will be sharply defined in the flat space limit.
\label{fig:inoutregions}}
\end{center}
\end{figure}

Now let us construct the S-Matrix.  Bulk time translations are generated by the dilatation operator $D$ in the CFT.  In section \ref{sec:States} we discussed the states that can serve as in and out states in the flat space limit, and from the CFT point of view, the most natural examples are the  time-regulated multi-trace primary states, which have zero total spatial momentum in the flat space limit.  Due to interactions, these are not eigenstates of the dilatation operator, but because $[D, K_i] = -K_i$\footnote{For any state $|n\>$, we have $K_i e^{i D t} |n\> = e^{i(D+1)t} K_i |n\>$, so $e^{i D t} | n \>$ is primary iff $|n\>$ is.}, primaries only mix with other primaries under time evolution.  As discussed in section \ref{sec:States}, our in and out states are only approximately primary, because we have regulated them in such a way that they are created or destroyed at fixed times. 
Although it seems miraculous from the CFT point of view, bulk locality guarantees that these states evolve trivially as free particles falling through AdS for until the particles meet and scatter at the center of AdS.

Using our definition of a $\Delta T$-regulated multi-trace primary state from equation (\ref{eqn:GeneralState}),  a preliminary formula for the S-Matrix can be written as the matrix element
\be
S_{g f} =  \< 0 | \mathcal{C}_g^\dag \ \! \mathcal{C}_f | 0 \>
\ee
The functions $f$ and $g$ create states of definite total energy and angular momentum in the flat space limit, and lead to a partial wave expansion of the S-Matrix.  As discussed in section \ref{sec:States}, we remove the singular components in the OPE expansion of the operators $\mathcal{C}_f$ and $\mathcal{C}_g$; this is just a generalization of the normal ordering procedure.  

Our formula is only preliminary because we must normalize the in and out states.  The unique disconnected $\mathbf 1$ component of the S-Matrix comes from the product of 2-pt correlators between operators $\CO_i$ and $\CO_i^\dag$ in $\mathcal{C}_f$ and $\mathcal{C}_f^\dag$, respectively, and it  allows us to normalize the in and out states.  We will see this explicitly for 2-particle states in the next section.  Once the states are normalized, we can compute the S-Matrix precisely.  We will see an example of this in section \ref{sec:4ptSMatrix}, where we use the normalization of 2-particle primary states to compute a simple 2-to-2 scattering amplitude.  We give a more general and formal discussion of perturbation theory in appendix \ref{sec:AppendixPerturbation}.

\subsection{Normalization of Primaries from Disconnected Correlators}
\label{sec:normalization}

The smearing functions derived in section \ref{sec:primaries} allow us to extract an in or out state using the operator $ \CO(x_1) \CO(x_2) $.  This in turn leads to a procedure for extracting a scattering amplitude from  the four-point functions $\< \CO(x_1) \CO(x_2) \CO(x_3) \CO(x_4)\>$,  since these contain within them $\< n,l,J| n',l',J'\>$.   In this section, we will first determine the appropriate normalization
factor on $f_{\omega,L}(t,\varphi)$ that leads to normalized two-particle states, and then we will  use the smearing functions to extract scattering amplitudes in the following section.   We should note, however, that although primary states are most natural from the CFT perspective, for computational purposes it is at least as easy and probably more familiar to compute scattering amplitudes in terms of plane waves, which we discussed in section \ref{sec:planewaves}.

We define our two particle state as 
\be
 \int d^d x_1 d^d x_2 \frac{f_{\omega,LJ}(x_1,x_2) }{N_{\omega, L J}}\CO(x_1)\CO(x_2) |0 \> &=&  | \omega,LJ\>_2.
\ee
From smearing the four-point function we can normalize these states  so that
\be
\int \prod_i dx_i \frac{f_{\omega,LJ}^*(x_1,x_2)  f_{\omega,LJ}(x_3,x_4)}{|N_{\omega, LJ}|^2}  \< \CO(x_1)\CO(x_2)\CO(x_3) \CO(x_4) \> = 1 
\ee
where we assume that the two smearing functions are separated by some time $T$.
To obtain the normalization, we consider the disconnected correlation function:
\be
\< \CO(x_1) \CO(x_2) \CO(x_3)\CO(x_4) \>_{disc.} &=&
\frac{1}{(x_{12}^2 x_{34}^2)^\Delta } + \frac{1}{(x_{13}^2 x_{24}^2)^\Delta }
+\frac{1}{(x_{14}^2 x_{23}^2)^\Delta }.
\ee
As we discussed in section \ref{sec:Parametrics}, we discard the first term on the RHS. The remaining two terms give equal contributions to the normalization, so we can just focus on one of them.  We will mainly be interested in the large $\omega$ limit, in which case the smearing functions simplify to
\be
f_{\omega, L, J}(x_1,x_2) &\approx& \exp \left( i \frac{\omega}{2}(t_1+t_2) - \frac{\omega}{8}(t_{12}^2 + \theta_{12\pi}^2 )  - \frac{t_1^2 + t_2^2}{\Delta T^2}\right) g_{LJ}(\hat x_1, \hat x_2),
\label{eq:simplesmear}
\ee
which one can check satisfy $K_i \CC_f =0$ in any dimension at large $\omega$. 
 In the above equation, $\theta_{12\pi} \equiv \Delta \theta(\hat x_1, \hat x_2) - \pi$, so that $\hat x_1$ and $\hat x_2$ are forced to be antipodal, and $g_{LJ}(\hat x_1, \hat x_2)$ is a function that guarantees that the total angular momentum quantum numbers of the state are $L$ and $J$. 
One can evaluate $\CA_{disc.}$ using contour integration, but an easier and less subtle prescription is to use the expressions for the 2-pt functions in terms of boundary wavefunctions, as in equation (\ref{eqn:OQuant}).  In the large energy limit, the normalization factors for these wavefunctions are 
\be
\frac{1}{\left(N^\CO_{E l} \right)^2 } \approx \frac{ 2 \pi^{d/2}}{\Gamma(\Delta) \Gamma\left(\Delta - \frac{d-2}{2} \right) } \left( \frac{E}{2} \right)^{2\Delta -d} ,
\ee
where $E = \Delta + 2n + l$.   We will compute $ {}_2\langle \omega, L | \omega', L' \rangle_2$ by writing the 2-point functions of $\CO$ as sums over these normalized wavefunctions, and to be specific we will compute the term $\frac{1}{(x_{13}^2 x_{24}^2)^\Delta }$.  This means we need to compute
\be
 \langle \omega, L J| \omega', L' J' \rangle & =& \frac{1}{|N_{\omega, LJ}|^2} \int dt_i d^{d-1} \hat x_i e^{-\frac{\omega}{8}(t_{12}^2 +  \theta_{12\pi}^2) - \frac{\omega'}{8}(t_{34}^2 + \theta_{34\pi}^2) - \frac{t_1^2 + t_2^2}{\Delta T^2} - \frac{(t_3-T)^2 + (t_4-T)^2}{\Delta T^2} } \nn \\ 
& & \times e^{i \frac{\omega}{2} (t_1 + t_2) - i \frac{\omega'}{2} (t_3 + t_4)}  g_{LJ}(\hat x_1, \hat x_2) g_{LJ}^*(\hat x_3, \hat x_4)  \\ 
& & \times \sum_{E_1 l_1 J_1, E_2, l_2, J_2} \frac{e^{i E_1 (t_1 - t_3)}}{\left( N^\CO_{E_1 l_1} \right)^2} \frac{e^{i E_2 (t_2 - t_4)}}{\left( N^\CO_{E_2 l_2} \right)^2} 
Y_{l_1 J_1}^*(\hat x_1) Y_{l_1 J_1} (\hat x_3) Y_{l_2 J_2}^*(\hat x_2) Y_{l_2 J_2} (\hat x_4) \nn
\ee
We have that $E_i = \Delta_i + 2n_i + l_i$, so changing the $n$ quantum number can only change the energy by an even integer (recall that these energies are in units of $1/R$), but changes in the angular momentum quantum number $l$ can change the energy by an even or odd integer.  The time regulator forces $t_1 - t_3$ and $t_2 - t_4$ to be near $T$, and so when $T \approx \pi m$ for some integer $m$, the sums over $E_i$ and $l_i$ will be coherent.  Otherwise there will be large cancellations among the terms, so we will henceforth assume that $T = m \pi$.  When the integer $m$ is even, the exponential factor will be independent of both the $n$ and $l$ quantum numbers, but when $m$ is odd they will be proportional to $(-1)^{l_i}$.  In the even case, the total elapsed time is a multiple of $2 \pi R$, so the particles from the initial state arrive back at their origin, and the sums over $l_i,J_i$ gives a product of spherical delta functions $\delta(\hat x_1, \hat x_3) \delta(\hat x_2, \hat x_4)$.  In the case of odd $m$, the angular momentum sum becomes
\be
\sum_{l_1 J_1, l_2, J_2}  (-1)^{l_1} (-1)^{l_2}Y_{l_1 J_1}^*(\hat x_1) Y_{l_1 J_1} (\hat x_3) Y_{l_2 J_2}^*(\hat x_2) Y_{l_2 J_2} (\hat x_4) = \delta(\hat x_1, -\hat x_3) \delta(\hat x_2, -\hat x_4)
\ee
The particles travel back and forth across AdS an odd number of times, so their initial and final states are at antipodal points on the boundary sphere.  The case $m=1$ will be the one that is useful for computing scattering amplitudes, because we want to give the particles one and only one chance to interact.
    
We can do the remaining two angular integrals to obtain a factor of order $\omega^{-\frac{d-1}{2}}$ from the Gaussian dependence on the angles in the large energy limit.  Thus we find
\be
\langle \omega, L J| \omega', L' J'\rangle = \delta_{L L'} \delta_{J J'}  \delta_{\Delta T}(\omega - \omega')\frac{\pi^{3+3d/2} \omega^{4 \Delta - 5d/2 } }{ 2^{8 \Delta -11d/2 - 15/2 } } 
\left( \frac{1} {\Gamma(\Delta) \Gamma \left( \Delta - \frac{d-2}{2}\right)} \right)^2 \nn
\ee
where the $\delta$ functions of energy are regulated by $\Delta T$.  Dropping the order one factors, this means that our two particle primary states are normalized when
\be
N_{\omega L J} \propto \omega^{2 \Delta - \frac{5d}{4} } 
\left( \frac{1} {\Gamma(\Delta) \Gamma \left( \Delta - \frac{d-2}{2}\right)} \right)
\ee

\subsection{The S-matrix from Smeared Correlation Function}
\label{sec:4ptSMatrix}

Now that we have fixed the normalization and functional form
of the smearing functions, we can use them to extract
scattering matrix elements from CFT correlation functions.
The quantity we want to calculate is the smeared four-point function:
\be
\CA_{conn} &=& \frac{1}{N_{\omega,L} N_{\omega',L'}} \int dt_i d\hat{x}_i f_{\omega, L}(t_1-t_{\rm in},t_2-t_{\rm in}, \hat{x}_1, \hat{x}_2)
f_{\omega',L'}^*(t_3-t_{\rm out}, t_4-t_{\rm out},\hat{x}_3, \hat{x}_4) \nn\\
&& \times  
\< \CO(x_1) \CO(x_2) \CO(x_3) \CO(x_4) \>,
\ee
where we will be using the large-dimension limit of the smearing functions from equation
(\ref{eq:simplesmear}).   In general, the four-point correlation function takes the form
\be
\< \CO(x_1) \CO(x_2) \CO(x_3) \CO(x_4) \> &=& \frac{1}{x_{13}^{2\Delta} x_{24}^{2 \Delta} } g(\sigma, \sigma', \rho),
\ee
where $\sigma, \sigma'$ and $\rho$ are conformal invariants 
\be
\sigma^2 &=& \frac{x_{13}^2 x_{24}^2 }{x_{12}^2 x_{34}^2 } \nn\\
\sigma'^2 &=& \frac{x_{14}^2 x_{23}^2 }{x_{12}^2 x_{34}^2 } \nn\\
\sinh^2 \rho &=& \frac{\det(x_{ij}^2)}{4 x_{13}^2 x_{24}^2 x_{12}^2 x_{34}^2 } .
\ee
Actually, $\rho$ is not independent of $\sigma, \sigma'$, but in fact
can be written in terms of them algebraically; however it will be
convenient to separate out its dependence explicitly.

The smearing functions extract a desired two-particle in state
from $\CO(x_1) \CO(x_2)$, and an out state from 
$\CO(x_3) \CO(x_4)$, where the shifts $t_{\rm in}, t_{\rm out}$ 
control the regions of time where the in and out states are prepared.
As discussed in section \ref{sec:hilbertspaces},
the in state will appear at time $t=t_{\rm in}$ to be near the boundary
of AdS, and will reach the center at a time of approximately $t\approx t_{\rm in} +\pi R/2$.
Similarly, the out state will be near the center of AdS at 
$t\approx t_{\rm out} - \pi R/2$.  Thus, to have non-trivial scattering,
we should choose our smearing regions as shown in Figure \ref{fig:inoutregions}.
An equivalent way of stating this constraint that does not make reference
to the behavior of the states outside of the flat-space region
$\rho \ll R$ is to say that the smearing functions with 
$t_{\rm in,out} = t_{\rm in,out}^{\rm (flat)}\pm \pi R/2$ prepare states that are
in the flat-space region at a time $t=t_{\rm in,out}^{\rm (flat)}$.
One must have $t_{\rm out} - t_{\rm in} \approx \pi R$ much larger than the thickness
$\Delta T$ of the time slices in the smearing functions in order for the free 
theory limit to be a good approximation when the scattering states are 
prepared.  Furthermore, to keep the entire scattering process
near the center of AdS, we will keep $t^{\rm (flat)}_{\rm in}, t^{\rm (flat)}_{\rm out} \ll 1$.
After the smearing functions extract the primary two-particle states,
the result is the transition element ${}^{\rm in}\< \omega, L| \omega',
L'\>^{\rm out}$.  We have chosen a normalization for
the primary states that approaches  $\< \omega, L,| \omega', L'\> = \delta(\omega-\omega') \delta_{LL'}  $ in the flat-space limit, so
in $d=2$ they are related to approximate two-particle plane waves $|p_1,p_2\>$ 
near the center of AdS by \cite{ECFT}
\be
| \omega, L ,J \> &\propto& \omega^{-1/2} \int d \hat{p} Y_{L}(\hat{p})
   \int d^2 q f(q) |  p + q, -p+q\>,
\ee
where $ f(q)$ is increasingly peaked around $q=0$ in the flat-space limit, 
and normalized by $\int d^2 q f^2(q)=1$. Consequently, we should find
\be
\CA_{conn} &= & {}^{\rm in}\< \omega, L| \omega', L'\>^{\rm out}
\propto \omega^{-1} \int d \theta e^{i L \theta}  T(s,t),
\label{eq:sphtoplane}
\ee
where $T(s,t)$ is the flat-space scattering matrix and $s,t$ are
Mandelstam variables, with $s=\omega$ and $t= (\omega^2 -4 m^2) \sin^2
\frac{\theta}{2}$. We will now check this explicitly in the special
case of quartic contact interactions for massless scalars, 
for which a simple limiting form of the correlation function was derived
in \cite{Gary:2009ae}.

It is convenient to change to a set of variables that are 
differences and sums of the operator insertion positions.
Specifically, we will use the separations that are constrained to
be small by the smearing functions:
\be
&& t_{12} = t_1 - t_2, \ \ \ \ t_{34} = t_3 - t_4 , \ \ \ \  \phi_{12} = \theta_1 - \theta_2-\pi, \ \ \ \ \theta_3 - \theta_4 -\pi, 
\ee
along with variables that are the sums and differences of coordinates
between the the in and out states:
\be
T = (t_1-t_{\rm in}) + (t_2-t_{\rm in}) +(t_3-t_{\rm out}) + (t_4-t_{\rm out}), \ \ \ \ && \Phi = \theta_1 + \theta_2 + \theta_3 + \theta_4, \nn\\
 t=\frac{ (t_1-t_{\rm in}) + (t_2-t_{\rm in}) }{2} - \frac{ (t_3-t_{\rm out})+(t_4+ t_{\rm out})}{2},\ \ \ \ &&
\phi = \frac{\theta_1 + \theta_2}{2} - \frac{\theta_3 + \theta_4}{2}.
\ee
The $\Phi$ and $T$ integrations may be done immediately, since by 
rotation and time-translation independence of the amplitude, all dependence
on them appears in the smearing functions.  The $T$ integration produces
an exponential factor $e^{-\Delta T^2(\omega-\omega')^2/16}$, which  at 
large $\omega \Delta T$ is just
a regularized delta function $\sim (\Delta T)^{-1} \delta(\omega-\omega')$.  
The $\Phi$ integration produces literally a
discrete delta function $\delta_{L,L'}$. Thus, without loss of generality
we set $\omega=\omega'$ and $L=L'$.
After the $T,\Phi$ integrations, the smeared correlation function is
\be
\CA_{conn} 
&\propto & \frac{1}{N^2_{\omega, L}}
 \int d t_{12} dt_{34} dt d\phi_{12} d\phi_{34} d \phi \frac{g(\sigma, \sigma',\rho)}{x_{13}^{2\Delta} x_{24}^{2\Delta}} 
\nn\\
&& \times \exp \left( -\frac{t_{12}^2 + t_{34}^2 +2 t^2}{2 \Delta T^2} + i L \phi + i t  \omega - \frac{\omega}{8} (t_{12}^2 + t_{34}^2 + \phi_{12}^2 + \phi_{34}^2 )\right)
. 
\ee
Now, for $\Delta T \ll R$, which is required in order to avoid the
scattering regions when we prepare our `in' and `out' states, 
we have $\rho \ll 1$, in which limit the correlation functions are singular.
Furthermore, the gaussian factors also enforce $\phi_{12}, \phi_{34} \ll 1$.
In this limit, the conformal invariants $\sigma, \sigma'$ can be
simplified: 
\be
\sigma &=& \cos^2 \left( \frac{\phi}{2} \right), \ \ \ \
\sigma' = 1-\sigma ,
\ee 
and four
point functions corresponding to contact interactions in AdS  take the simple form \cite{Gary:2009ae, holographyCFT}
\be
g(\sigma, \sigma', \rho) 
&\propto& \frac{\CM(\sigma) \sigma^{1-k} (1-\sigma)^{2\Delta +k -2} }{ \rho^{4 \Delta + 2k - 3}},
\ee
where $\CM(\sigma)$ is the angular-dependence of the flat-space scattering amplitude $T(s,t)$:\footnote{Note that quartic contact terms with four
identical scalars give scattering amplitudes symmetric in $\sigma 
\leftrightarrow \sigma'$.}
\be
\CM(\sigma) &\equiv & T(s=1, t=-\sigma').
\ee
The parameter $k$ is the number of derivatives in the contact interaction. 
Recall that the scattering angle $\theta$ is related to $t$ by $-t/s = \sin^2(\theta/2)$, so our integration variable $\phi$ therefore enters
in $T(1,-\sigma')$ exactly as the scattering angle of the flat-space
scattering process!
 
The limit of $\rho$ in the denominator is very singular near the saddle
point $\phi_{12} , \phi_{34} \ll 1$ and requires more care.  First,
note that for $t,t_{12},t_{34} \ll 1$, 
\be
\rho^2 &\approx& \left[ \frac{ 4 t \cos {\phi_{12} \over 2} \cos {\phi_{34} \over 2} \sin \phi - t_{34} \left(  \sin \phi_{12} + 2 \cos \phi \cos {\phi_{12} \over 2} \sin {\phi_{34} \over 2} \right) - t_{12} \left( 2 \cos \phi \cos {\phi_{34} \over 2} \sin {\phi_{12} \over 2} + \sin \phi_{34} \right) }{8 \cos^2 {\phi \over 2}} \right]^2  \nn\\
\ee
and depends only on $t,t_{12}, t_{34}$ through the linear combination $\tau$,
defined so that $\rho^2 \approx \tau^2/[\cos^2 {\phi \over 2} ]^2$.
At this point, the exponential is most sensitive to the phase terms,
which are linear in $t_{ij}, \phi_{ij}$.  This can be seen by
rescaling them by a factor of $\omega$, i.e. $t_{ij} = {t'_{ij} \over \omega},
\phi_{ij} = {\phi'_{ij} \over \omega}, \tau={ \tau' \sin \phi \over 2 \omega} $, and taking the large
$\omega, \Delta T$ limit:
\be
\CA_{conn} 
&\propto& \frac{\omega^{4\Delta +2k -8}e^{-i \omega \pi} }{
N_{\omega,L}^2 }  \int
 d\tau' d\phi dt'_{12} dt'_{34} d\phi'_{12} d \phi'_{34} \frac{\CM(\sigma) \csc \phi }{(\tau'^2)^{2\Delta + k -3/2}}  \\
&& \times
\exp\left( i L \phi - i \tau 
 - \frac{ t_{12}'^2 +t_{34}'^2 + \phi_{12}'^2+\phi_{34}'^2 
 - 2i ((t'_{12}\phi'_{12} + t'_{34} \phi'_{34}) \cot \phi 
 +   t'_{34} \phi'_{12}  + t'_{12} \phi'_{34})  }{8 \omega} \right) . \nn
\ee
Now, the $t_{12}',t_{34}',\phi_{12}',\phi_{34}'$ integrations may easily
be done, leaving only
\be
\CA_{conn} 
 &\propto&  \frac{\omega^{4\Delta +2k -6} e^{-i \omega \pi}}{
N_{\omega,L}^2 } 
\int
 d\tau' d\phi \frac{\CM(\sigma) }{(\tau'^2)^{2\Delta + k -3/2}} 
e^{i L \phi  - i \tau'} 
  \propto  \frac{\omega^{4\Delta  -6} }{
N_{\omega,L}^2 }   \int  d\phi T(s,t) e^{i L \phi }. 
\ee
where we have used the fact that $s^{2k} T(1,-\sin^2 (\phi/2)) = T(s,t)$
for a contact interaction with $2k$ derivatives, and $\omega=s$ by 
construction. 
After including
the normalization of the states, and the  factor of $\omega$ in
eq. (\ref{eq:sphtoplane}) from changing basis
to plane waves, this gives the flat-space amplitude.

\sect{Computations of IR Effects}

In the construction of section \ref{sec:States}, AdS plays the role of a particularly symmetric and versatile `box'.  The negative curvature of AdS means that a space-like sphere of radius $r$ will have surface area proportional to $e^{r/R}$, so that by Gauss's law, the field created by a point charge will fall off exponentially, as though the field has a mass of order $1/R$.  This means that AdS automatically regulates long distance effects and IR divergences, precluding the possibility that particles can interact while they are far outside our flat `universe' at the center of AdS.  In this section we will confirm these intuitive expectations with concrete calculations.

\subsection{Decoupling of Long Distance Effects -- Computation of Massless Exchange}

One might wonder if particles can interact while they are far from the center of AdS.  If this were possible, it would jeopardize our construction of the S-Matrix, since it would mean that particles could interact before they enter our universe.  Physically, we expect that this cannot occur because the potential created by one particle will fall off exponentially at large distances, making it negligible at the location of the second particle.  To directly confirm this reasoning, we will examine the contribution of the t-channel exchange of a massless particle to the s-wave $2 \to 2$ scattering amplitude.

We begin with a theory of massless $\phi$ and $\chi$ fields with a cubic interaction $\phi^2 \chi$.  At second order in the interaction we find an amplitude
\be
{\cal M} = \int d^{d+1}x d^{d+1}y \sqrt{g_x g_y}  \langle n , 0  | T [ \phi(x)^2 \chi(x) \phi(y)^2 \chi(y) ] | n, 0 \rangle .
\ee
This includes s, t, and u channel contributions.  To compute the t-channel (or u-channel) terms, we need to use the wavefunction
\be
\psi(x,y) =  \langle 0 | \phi(x) \phi(y) | n,0 \rangle
\ee
in AdS.  Contracting the $\chi$ fields into a massless propagator gives
\be
{\cal M}_t = \int d^{d+1}x d^{d+1}y \sqrt{g_x g_y} \ \! \psi(x,y)^\dag \psi(x,y) 
e^{-\Delta_\chi \sigma(x,y)} F \left( \Delta_\chi, \frac{d}{2}, \Delta_\chi + 1 - \frac{d}{2}, e^{-2 \sigma} \right) ,
\ee
where $\Delta_\chi = d$ because $\chi$ is massless.  

We will now show on very general grounds that this t-channel exchange amplitude becomes exponentially small at large distances.  To clarify the physics of the computation, it's useful to rewrite the AdS metric
using geodesic distance from the center of AdS, $\kappa$, in place of $\rho$.  This is
\be
ds^2 = - \cosh(\kappa)^2 dt^2 + d\kappa^2 + \sinh(\kappa)^2 d \Omega^2 ,
\ee
where we have that $\sec(\rho) = \cosh(\kappa)$ and $\tan(\rho) = \sinh(\kappa)$.  To set up the calculation, we will need to use the fact that
\be
\langle 0 | \phi(x) \phi(x') | n,0 \rangle_2 = \left( \frac{1}{ \cosh(\kappa) \cosh(\kappa')} \right)^{\Delta + n} \sum_{i,j}^n c_{ij} w^i \sigma(x,x')^j ,
\ee
where $\Delta$ is the dimension of the operator dual to $\phi$ (eg it is $d$ for a massless scalar $\phi$) and
\be
w = \cosh(\log(y)) \ \ \ \mathrm{with} \ \ \ y = e^{\tau - \tau'}  \frac{\cosh(\kappa')}{\cosh(\kappa)} .
\ee
This follows from conformal symmetry, and the fact that $|n,0 \rangle$ is a primary state \cite{ECFT}.   Now since we are interested in the IR limit, we will be taking the region where $\kappa, \kappa' \gg 1$.  This means that generally speaking, we can approximate $\cosh(\kappa') \approx e^{\kappa}$, and that we can ignore power law terms in $\kappa$ compared to exponentials.  In this case we can simplify
\be
w \approx \cosh(\tau - \tau' - (\kappa - \kappa')) ,
\ee 
and we can ignore the power-law dependence on $\sigma$ and the hypergeometric function.  We will just look at one of the $c_{ij}$ terms, since each vanishes individually in the IR limit.
\be
{\cal M}_t & \approx & \int d^{d+1}x d^{d+1}x' 
\frac{e^{(d - 2 \Delta - 2 n)(\kappa + \kappa')} \cosh(\tau - \tau' - (\kappa - \kappa'))^{i+i'} }{\left( e^{\kappa + \kappa'} (\cosh(\tau - \tau') - \cos(\theta - \theta')) + \cosh(\kappa - \kappa') \cosh(\tau - \tau') \right)^{\Delta_\chi} } \nonumber \\
& \lesssim & \int d^{d+1}x d^{d+1}x' e^{(\kappa + \kappa') (d - 2 \Delta - \Delta_\chi)} \to \mathrm{finite} ,
\ee
since we know that $\Delta \geq d$.  We see from the explicit expression that as expected, the contribution from large $\kappa, \kappa'$ falls off exponentially, so particles cannot interact before they enter our universe at the center of AdS.

\subsection{Soft and Collinear Emission}

One might be concerned about collinear effects spoiling the flat space S-Matrix, since nearly collinear fields might continue to interact far outside our universe, to which scattering is supposed to be confined.  In fact, AdS space acts as an elegant regulator for soft and collinear divergences in quantum field theory, and there are no soft or collinear interactions over distances larger than $R$.  There are two related effects: first, as we discussed above, the curvature of AdS means that even massless fields fall off exponentially at distances larger than $R$.  Second, the spectrum of quantum field theory in AdS is discrete, so the phase space integrals we would encounter in flat space turn into sums in AdS, regulated by a minimum energy $1/R$.  This second effect also guarantees that loop integrals will be regulated in the Infrared by AdS space.

Any theory with a discrete spectrum will be IR finite -- for instance, one could regulate quantum field theory by putting the universe in a finite sized box.  So let us explore how the curvature affects soft and collinear emission.  Consider a theory of massless scalar fields with a $\phi^2 \pi$ interaction.  The amplitude for soft and collinear emission in flat space is
\be
\frac{1}{p \cdot k_s } \approx \int d^d x \frac{1}{x^{d-2}} \langle 0 | \phi(x) \pi(x) | p, k_{s} \rangle ,
\ee
where the soft particle is a massless scalar with momentum $k_s$, and $p$ is the momentum of a hard $\phi$ particle.  This amplitude can be understood in a variety of ways, but perhaps the simplest and most physically transparent computation views the soft particle as a background in which the hard particle propagates.  Since this $\pi$ background is a small perturbation, the amplitude is given by the phase shift of the $\phi$ particle, which one can compute by an integral along its world-line, giving
\be
e^{i \int d \tau \pi(x(\tau))} - 1 \approx i \int_0^\infty d \tau e^{i \tau \hat p \cdot k_s} = \frac{1}{\hat p \cdot k_s} ,
\ee
as expected.  We can perform the equivalent calculation in AdS, given by
\be
\int d t d \rho d^{d-1} \hat x \ \! \frac{ \sin^{d-1} (\rho) }{ \cos^{d+1} (\rho) } \frac{1}{ (\cos \rho - \cos t)^{\Delta_\phi} } \langle 0 | \phi(x) \pi(x) | p_{\phi}, p_{\pi s } \rangle ,
\ee
either via a direct computation or using an integral along the world-line of the $\phi$ particle in a background given by the $\pi(x)$ wavefunction in AdS.  The world-line calculation gives
\be
\exp \left[ i \int  \frac{d \tau}{\cos \tau} \pi(\rho(\tau), t(\tau)) \right]  - 1 \approx i \int  \frac{d \tau}{\cos \tau} \pi(\rho(\tau), t(\tau)) ,
\ee
and so we need to evaluate the $\pi_{k_s}(\rho, t)$, but these are wavefunctions that we determined in section \ref{sec:States}.
A massless $\phi$ particle propagates in the AdS metric with $t(\tau) = \tau$ and $\rho(\tau) = \tau$.   Since our $\phi$ particle propagates from the center of AdS to the boundary, we can simplify the calculation by looking at the $l = 0$ wavefunction for $\pi$.  For the soft factor, we find
\be
i \int_0^{ \frac{\pi}{2} }  \frac{d \tau}{\cos \tau} \pi(\tau, \tau) = \frac{i}{N_{n00}^\pi} \int_0^{\frac{\pi}{2}}  \frac{d \tau}{\cos \tau} e^{i E_{n,0} \tau} \left[ \cos^{d} \tau \ \! F \left(-n, d+n, \frac{d}{2}, \sin^2 \tau \right) \right] .
\ee
This integral converges for all physical values of $n, l,$ and $d$, so AdS has regulated the soft singularity corresponding to $k_s \to 0$ in flat space.  If we take the limit of large $n$ with the other parameters held fixed, the integral is proportional to $1/n$ as expected -- in this limit we reproduce the usual result from flat space, namely that the soft emission amplitude is proportional to $\frac{1}{\hat p \cdot k_s}$.

\subsection{Multiple Scattering}
\label{sec:multiplescat}

We have constructed our in and out states in terms of healthy normalizable modes in AdS, which correspond to genuine states in the CFT.  However, a pure normalizable mode in AdS is an eternal standing wave.  If we use these modes to make localized wave packets that correspond to particles, then they will oscillate back and forth in AdS forever, passing through our flat universe at the center of AdS an infinite number of times.   If we have several particles, then they can scatter every time, leading to an infinite number of interactions!

In section \ref{sec:States} we avoided this problem by regulating our states -- roughly speaking, we inserted the particles and then removed them after one scattering event.  However, it would be interesting to see explicitly how multiple scatterings arise when we remove the time regulator, and that is the goal of this subsection.  Multiple scattering will be invisible at lowest order in perturbation theory, since by definition it requires at least two scattering events, so we will need to examine either s-channel exchange at tree level with a cubic interaction, or a one-loop effect with a quartic interaction.

Consider $\lambda \phi^4$ theory at one-loop.  If we drop our regulator on the initial and final time intervals for the in and out states, the scattering amplitude becomes
\be
\CA &=& \int d^{d+1} x d^{d+1} x' \sqrt{g g'} {}_2\< n,0 | \lambda \phi^4(x) \lambda \phi^4(x') | n, 0 \>_2 \\
&=& \lambda^2 \int_{-\Delta t }^{\Delta t}  d^{d+1} x d^{d+1} x' \frac{sin^{d-1} \rho \sin^{d-1} \rho'}{\cos^{d+1} \rho \cos^{d+1} \rho'}  \left( \cos \rho e^{it} \right)^{2\Delta + 2n}
\left( \cos \rho' e^{-i t'} \right)^{2\Delta + 2n} \left( G_{\Delta_e}(x,x' ) \right)^2, \nn
\ee
where $G_{\Delta_e} (x,x') $ is the scalar propagator for the exchanged scalars, which have a dimension $\Delta_e$ allowed to be different from the dimension $\Delta$ of the external $\phi$'s:
\be
G_{\Delta_e} (x,x') &=& e^{-\Delta_e \sigma} F(\Delta_e, d/2, \Delta_e +1 -d/2, e^{-2 \sigma}) \nn\\
 &\approx& \frac{e^{-\Delta_e \sigma} }{(1-e^{-2\sigma})^{d-1}} \ \ \ (\mathrm{Re} [\sigma ] \ll 1).
\ee
The crucial point is that the propagator is periodic up to a phase when $\sigma \to \sigma +  i \pi m$ for integral $m$.  In particular, there is a singularity in the propagator at $\sigma =  i \pi m$.  These singularities occur when $x$ and $x'$ differ by a multiple of a half period of oscillation in AdS, and are the multiple scattering events.  At large $n,n'$, the external wavefunctions only have support at very small values of $\rho, \rho'$, so we can take $\rho, \rho' \ll 1$, and the geodesic distance between $x$ and $x'$ to be the flat space distance $\sigma^2 \approx (\vec x - \vec x')^2 - (t-t')^2$.  But this reduces our calculation to that of the flat space loop integral, with the caveat that the loop integral gets repeated for each scattering event.

We have seen that without a regulator on the times of the initial and final states, we have multiple scattering in AdS, and that for the simple case of $\phi^4$ theory at $\CO(\lambda^2)$ we obtain a multiple of the one-loop amplitude.

\section{Discussion}

We have shown that time-regulated multi-trace primary states in a large N CFT form a basis for the in and out states in the flat space limit of its AdS dual.  While these states are particularly natural from the point of view of the CFT, we also used CFT operators to construct states that become the more familiar plane waves in the flat space limit.  Using either basis, one can compute the flat space S-Matrix to arbitrary precision in terms of CFT data, such as the matrix elements of CFT states or CFT correlation functions.  Since one can create a black hole using well-separated but highly boosted initial states, this means that with adequate CFT data, our construction would allow us to extract scattering amplitudes with black holes as intermediate states.  Black holes seem to be generic gravitational objects, so it would be fascinating to understand what features of CFTs give rise to the Hawking evaporation of sub-AdS scale black holes, which turn into standard flat space black holes in the flat space limit.

After performing a sample computation in terms of the primary states, we also examined infrared effects in order to argue that interactions outside of the center of AdS do not pollute the S-Matrix in the flat space limit.  We showed that interactions between particles near the boundary of AdS are completely negligible, that standard soft and collinear limits obtain, and that multiple scatterings over times of order the AdS length scale are absent for the states that we consider.  All of these effects can be understood by thinking of AdS as a particularly elegant `box' \cite{Callan:1989em} that regulates IR divergences while maintaining the same number of symmetries as in flat space.  As particles travel away from the center of AdS, they become exponentially better separated then their flat space equivalents, so in a sense, they quickly become IR free.

The flat space S-Matrix has been obtained as a limit of AdS correlation functions before \cite{Polchinski:1999ry,Susskind:1998vk,Giddings:1999jq}, and the physics of this limit was very nicely explained early on in \cite{Polchinski:1999ry,Susskind:1998vk}.  However, recently there has been some controversy over whether the S-Matrix can be extracted to arbitrary precision from CFT correlators due to concerns about singularities and the limitations of wavepackets.  By showing directly which states in the CFT correspond to AdS configurations consisting of well-separated incoming and outgoing particles, we have argued that these concerns are unwarranted and that the flat space S-Matrix can be computed exactly from CFT data.  

Our examination of in and out states was also motivated by a desire to have a precise and natural definition of the S-Matrix from the CFT point of view in order to make contact with a very interesting recent formula of Penedones \cite{mellin}, which computes the flat space S-Matrix as the limit of the Mellin transform of CFT correlation functions.  Although very few CFT correlation functions have been computed, a host of scattering amplitudes are known, perhaps because flat space tree amplitudes are simply rational functions in momentum space.  The Mellin transform of CFT correlators seems to give expressions that are very similar to momentum space in the flat space limit, so there is hope that it could yield methods for computing many more complicated CFT correlation functions.  Conversely, it may also be possible to obtain information about CFTs using the flat space S-Matrix.

The existence of a natural CFT formula for the flat space S-Matrix suggests that it should be possible to connect these two objects in a more fundamental way, by associating the unitarity and analyticity of the S-Matrix with equivalent properties of matrix elements in a CFT.   A direct connection could only emerge in the flat space limit, but it might not require any intermediate regulator acting on the CFT states, such as the time regulator we introduced when defining in and out states via equation (\ref{eqn:GeneralState}).  We hope to explore this CFT/S-Matrix connection in future work.

\subsection*{Acknowledgments}

We would like to thank Nima Arkani-Hamed, David Berenstein, Andy Cohen, Steve Giddings, Shamit Kachru, Jo\~ao Penedones, and Joe Polchinski for interesting discussions, and Ami Katz for early collaboration and comments on the draft.  We would also like to thank the Institute for Advanced Study and the Aspen Center for Physics for hospitality.  JK is supported by SLAC, which is operated by Stanford University for the US Department of Energy under contract DE-AC02-76SF00515.  ALF is supported by DOE grant DE-FG02-01ER-40676 and NSF CAREER grant PHY-0645456.

\appendix

\section{A Comparison of Formal Definitions of the S-Matrix}
\label{sec:AppendixLSZ}

In this appendix we will briefly review the physical intuition that allows for a precise definition of the S-Matrix via the LSZ prescription.  Then we will discuss the modifications necessary to define an S-Matrix in AdS space, and see how these modifications disappear in the flat space limit.

The S-Matrix connects asymptotic `in' and `out' states consisting of particles that are created and absorbed in the infinite past and future.  There are two steps required to give a formal definition of the S-Matrix:  (1) we need to define single particles in a non-perturbative way, and (2) we must construct the in and out states where these particles propagate freely in the infinite past and future.  

In flat space, the first step can be carried out for massive particles, since there is a unique lowest energy state corresponding to a single massive particle at rest.  However, for massless particles this step is not obvious, and indeed, we know that massless particles tend to cause IR divergences which require us to define our observables in terms of jets with an experimental resolution.  In AdS massive and massless particles can be defined because the spectrum in AdS is discrete, but this feature is not particular to AdS -- we would find the same result if we put our theory in a very large box, or used another physical IR regulator. 
In both flat space and AdS the formal definition of particles is not completely satisfactory, for example the proton is not the unique state with its conserved quantum numbers, and so if it can decay, formally speaking it is not a particle from the S-Matrix point of view.  Clearly in both flat space and AdS/CFT, the lowest energy (dimension) state with a given set of conserved quantum numbers can be taken as a well-defined particle at rest. However, if a `particle' such as the proton turns out to be stable, it would be more difficult to give it a formal definition, because it overlaps with the continuum of states involving a positron and many other particles.

For the second step, we need to define multi-particle states with the property that the particles are guaranteed to propagate freely in the asymptotic past.  In flat space this is accomplished by setting up wavepackets where the particles do not overlap in momentum space, meaning that we can be certain that in the limit of large times, the particles are very far apart.  In AdS, particles can become very well-separated as they move towards the boundary, but for finite AdS length $R$, they can only propagate backwards in time for a finite time before they reach their closest approach to the boundary and begin to fall back towards the center of AdS.   As we take the flat space limit $R \to \infty$ the particles can propagate for an infinite time, and we recover the flat space prescription, but to better understand our S-Matrix construction at finite $R$, let us review how the LSZ prescription changes if our states can only propagate into the past and future for a very large but finite time.

To understand this, we merely have to make a slight modification to a  textbook proof of the LSZ reduction formula. To begin, one assumes that the interacting theory has a vacuum $|0\>$,  and exact one-particle states $|k\>$, obtained by boosting particles at rest.  The CFT analog is a single-trace primary state (a particle at rest) and its descendants (boosted versions with different momenta).  In the flat space limit $N \to \infty$ and the notion of a single-trace primary becomes exact.  In the LSZ prescription in flat space, one takes a bulk field $\phi(x)$ with some overlap with single-particle states $|k\>$, one subtracts any vev $\< \phi \>$ and rescales by $\phi \rightarrow \phi /\sqrt{Z}$, $\sqrt{Z} = \< k | \phi(0) | 0 \>$, so that
\be
\< k| \phi(x) | 0\> &=& \phi_k(x).
\ee
In flat space, $\phi_k(x) = e^{i k x}$ by Lorentz invariance.  A general single-particle wavepacket state is
\be
|f \> &=& \sum_k \frac{1}{2 \omega_k} F_k | k \> , \nn\\
f(x) &=& \sum_k \phi_k^*(x) F_k .
\ee
Here, $F_k$ are wavepacket mode coefficients of our choosing.
One defines the smeared field $\phi^f$ as
\be
\phi^f(t) &\equiv& i \int d^d x ( \phi \dot{f} - f \dot{\phi}) .
\ee
This has the desired property that for a plane wave $f(x) =e^{i p \cdot x}$ 
the creation operators
$a_k^\dagger$ are extracted with coefficient 1, and the annihilation
operators are canceled.  Thus $\phi^f$ creates one-particles
states (as well as multi-particle states), but does not annihilate
one-particle states:
\be
&& \< 0 | \phi^f(t) |0\> = 0 , \ \ \ \ 
\< 0 | \phi^f(t) | k \> = 0, \ \ \ \
\< k | \phi^f(t) | 0\> = F_k,
\ee
A key point is that as $t$ is taken to the infinite past
or future, all contributions from multiparticle states are suppressed
by time oscillation factors.  One then has a precise way to 
make two-particle states - simply take two $\phi^f$'s and look in the far
past or future:
\be
\lim_{t\rightarrow \pm \infty} \phi^{f_1} \phi^{f_2} | 0\> &=& |f_1, f_2\>.
\label{eq:twopartlszstates}
\ee
In order for these to be strictly two-particle states, it is important
that  $f_1,f_2$ have no overlap in momentum-space,
so that the two particles are really separated in the infinite past
(components where both particles have the same momentum will stay close 
to each other at all times).  It is also clear that due to the
presence of multi-particle states in $\phi^f$ at finite times, 
$\phi^f | 0\>$ is not an eigenstate of the Hamiltonian in general
and so neither is $\phi^{f_1} \phi^{f_2} | 0 \>$.  This means that although time propagation is trivial in the infinite past, the S-Matrix can be very non-trivial.

 The LSZ formula states that scattering amplitudes of
arbitrary wavepacket states may be extracted from the time-ordered
Green functions. For two-to-two scattering, it takes the form
\be
{}^{\rm out}\< f_3, f_4 | (S-1) | f_1, f_2 \>^{\rm in} &=&
(i)^4 \int d^d x_1 \dots d^d x_4 f_1(x_1) \dots f_4^*(x_4)
\Pi_{i=1}^4 (\Box_i - m_i^2) \< 0 | T\left\{ \phi_1 \dots \phi_4 \right\} |0\> .
\nn\\
\label{eq:lsz}
\ee
If we only have a finite window for scattering, we make the slight modification that 
the time integrals are limited to begin at some past time 
$-T$ and to proceed to a future time $T$.
Now, $(\Box -m^2)f_i =0$, and $f$ is spatially
localized, so one has
\be
i \int_{-T}^{T} dt \int d^d x f(x) (\Box - m^2 ) A(x) &=&
  i  \int_{-T}^{T} dt \int d^d x \left(
-f \d_t^2 A + A(\nabla^2 - \mu^2)f \right) \nn\\
  &=&  - i \int_{-T}^{T} dt \d_t \int d^d x \left( f \dot{A} - \dot{f} A\right) \nn\\
  &=& \left[ \lim_{t\rightarrow T} - \lim_{t\rightarrow -T} \right] A^f(t) .
\ee
Thus, the LSZ formula for the S-matrix gives
\be
(S-1) &=& \left( \Pi_{i=1}^4 \left[ \lim_{t_i\rightarrow T} - \lim_{t_i\rightarrow -T} \right] \right)
\< 0 | T\left\{ \phi^{f_1} (t_1) \dots \phi^{f^*_4} (t_4) \right\} | 0 \> \nn\\
&\approx &  \left( \Pi_{i=1}^2 \left[ \lim_{t_i\rightarrow T} - \lim_{t_i\rightarrow -T} \right] \right)
\< 0 | T\left\{ \phi^{f_3^*} (t_3) \phi^{f^*_4} (t_4) \right\} | f_1, f_2 \>^{\rm in} \nn\\
&\approx& {}^{\rm out} \< f_3, f_4 | f_1, f_2 \>^{\rm in} - 
{}^{\rm in}\< f_3, f_4| f_1, f_2 \>^{\rm in} ,
\ee
which does indeed reproduce the $(S-1)$ matrix, up to contributions
from additional particles in the initial and final states, due to the fact
that oscillations  kill them completely only in the $T \rightarrow 
\infty$ limit.  Also, it is manifest here that our `in' and `out' states
live at finite future and past times, corresponding to $T$.  
In practice one can calculate the right-hand side
of eq. \ref{eq:lsz} for finite $T$, in order to see how well
scattering of the states $\phi^{f_1}(\pm T)\phi^{f_2}(\pm T)| 0\>$
approximates the exact S-matrix.  Consider $\lambda \phi^4$ theory.
In this case, at $\CO(\lambda)$, a single insertion of the interaction
vertex $\lambda \int d^{d+1} w \phi^4(w)$ just gives four propagators 
from the $\phi(x_i)$ contractions, which are turned into delta functions
by the $(\Box - m^2)$ factors.  Performing the $t_i,x_i$ integrations, one
obtains
\be
(S-1) &=& \lambda \int_{-T}^{T} d w_0 \int d^d w f_1(w) f_2(w) f_3^*(w)
f_4^*(w) .
\ee
If we choose plane waves $e^{i p_i \cdot x}$ for the wavefunctions
$f_i$, then we get the result
\be
(S-1) &=& \lambda \frac{\sin (T \sum_i p_i^0)}{\sum_i p_i^0}
\delta^d( \sum_i \vec{p}_i ),
\label{eq:finitetime}
\ee
which turns into an energy conserving delta function in the limit $T \to \infty$.

\section{Perturbation Theory  of the S-Matrix in AdS/CFT}
\label{sec:AppendixPerturbation}

In the body of the text we used the correspondence between free scalar fields and CFT operators to construct in and out scattering states appropriate for the flat space limit of AdS.  Then we argued that these states can be generalized in a natural way to a full non-perturbative CFT in the limit of $N, \lambda \to \infty$ as long as we examine states with energy $ER \to \infty$.  The goal of this appendix is to elucidate these arguments by showing how one could setup perturbation theory, which lies between the free theory and the full non-perturbative CFT.

We begin by imagining that we understand the CFT order by order in $1/N$, so we have an expansion for the exact primary operator $\CO_N$ which looks like
\be
\CO_N(x) = Z_N \CO_\infty(x) + \frac{1}{N} \int d^d y \ \! g(y) \CO_\infty(x- y) \CO_\infty(y) + ...
\ee
and there will also be $1/N$ corrections that mix different primary operators which we ignore for brevity.  This operator has dimension
\be
\Delta_N = \Delta_\infty + \frac{1}{N} \Delta_1 + \frac{1}{N^2} \Delta_2 + ...
\ee
$\CO_N$ is a primary under the full interacting conformal group, although at finite $N$ it cannot be said to be single-trace, as that concept is not well-defined.  We can assume that we know the exact dimension $\Delta_N$, in the same way that in flat space we know the physical mass of particles such as the electron.  Then, at each order in perturbation theory we will correct for the interactions by adding the counter-terms $\Delta_1$, $\Delta_2$, etc so that the operator $\CO_N(x)$ will have the correct physical dimension to that order in perturbation theory.  Similarly, the normalization of $\CO_\infty$ will be shifted by interactions, so we have introduced the factor $Z_N$ as a field strength renormalization to insure that $\CO_N$ can create normalized single-particle states.

Now we construct scattering states out of $\CO_N$ using equation \ref{eqn:GeneralState}.  The perturbative version of the conformal algebra will act nicely on $\CO_N$ if we take the algebra and the operator to be at the same order.  This guarantees that the scattering states will be time-regulated primary states consisting of particles with the correct renormalized mass and total energy (dimension) given by the smearing function.  However, the correlation functions of $\CO_N$ will be non-trivial and scattering processes will occur because of the non-linear terms in the expansion of $\CO_N$ in $\CO_\infty$.  Of course to do these computations via an explicit perturbation series, we would most likely use the AdS/CFT correspondence and a bulk Lagrangian to determine $\CO_N$ in terms of the free operator $\CO_\infty$.

\section{Extracting Coefficients from Correlators}
\label{sec:AppendixNorms}

In this appendix, we will derive the Laurent expansion  of primary
operators using the
2-point correlation functions.  First, as a warm-up, we will show
how this is done in two dimensions, where the factorization
into left-movers and right-movers allows one to avoid dealing with
angular dependence in the two-point function. The analysis in higher
dimensions is more involved because the spherical harmonics are more
complicated, but is not conceptually different.  In general, one 
can derive the Laurent expansion by writing the 2-point correlator
in two different ways.  First, use the definition of the
Laurent expansion inside the correlator:
\be
\< \CO(\tau) \CO(0) \> &=& \sum_{n,m} N_n(h) e^{-\tau (h+n)} N_m(h)
  \delta_{nm} = \sum_n N_n^2(h) e^{-\tau (h+n)},
\ee
where we have just included the left-moving sector, since the
left- and right-moving sector factorize.  We may also
use the explicit form of the 2-point correlators:
\be
\< \CO(\tau) \CO(0) \> &=&  \frac{e^{\tau h}}{(e^\tau -1)^{2h}}
  =  \frac{e^{-\tau h}}{(1-e^{-\tau})^{2h}} 
   = \sum_m e^{-\tau(h+m)} \frac{\Gamma(2h+m)}{\Gamma(2h)m!} .
\ee
By matching powers of $e^{-\tau}$, we can read off that
\be
N_n(h)^2 &=& \frac{\Gamma(2h+m)}{\Gamma(2h) m!}.
\ee

Now, let us turn to higher dimensions.  The conformal group no longer
reduces to left-moving and right-moving pieces, so we have to
take into account angular-dependence in the correlator.  Let us
again start by using the Laurent expansion in the 2-point function:
\be
\< \CO(\hat{x}', \tau) \CO(\hat{x}, 0) \> &=&
\sum_{n,l,J} N_{n,l}^2(\Delta) e^{-\tau (\Delta + 2n +l)}
Y_{lJ}^*( \hat{x}') Y_{lJ}(\hat{x})  .
\ee
As in the body of the text, $\hat{x}$ denotes the angular variables in
the embedding space (i.e. $\hat{x} \cdot \hat{x} =1$).
We have used that there is no $J$-dependence in the Laurent
coefficients.  
We can do the sum over $J$ by using the fact that 
$\sum_J Y_{lJ}^*(\hat{x}') Y_{lJ}(\hat{x})$ can depend only on $\cos \gamma =
\hat{x}' \cdot \hat{x}$, and its dependence is set through the Gegenbauer polynomials $C_l^{(d-2)/2}$, which are generalizations of Legendre polynomials.  
In general dimension, these are 
determined up to an overall normalization 
by the orthogonality condition $\int d \gamma 
\sqrt{g} C_l(\cos \gamma) C_{l'}(\cos \gamma)  =
\int d \mu (1-\mu^2)^{(d-3)/2} C_l(\mu) C_{l'}(\mu) \propto \delta_{l l'}$.
The conventional notation for such polynomials is
$C_l^\lambda (\mu)$, $\lambda = (d-2)/2$, which satisfy the following
important properties.  First, there is the normalization:
\be
\int_{-1}^{1} (1- \mu^2)^{\lambda - \half} C_l^\lambda(\mu) C_{l'}^\lambda(\mu) d\mu
 &=& \delta_{l l'} \frac{ (2\pi) \Gamma(2\lambda +l)}
{ 2^{2\lambda} \Gamma^2(\lambda) l! (l+\lambda) } .
\ee
It is also helpful to know their expression in terms of hypergeometric
functions:
\be
C_l^\lambda(\mu) &=&
\frac{\Gamma(2\lambda+l )}{\Gamma(2\lambda)l!} F(2\lambda +l, -l, \lambda+\half,
\frac{1-\mu}{2}) .
\ee
From this, one can immediately read off their values when $\mu=1$:
\be
C_l^\lambda(1) &=& {d + l -3 \choose l}.
\ee
The sum over $J$ of spherical harmonics is then fixed to be
\be
\sum_J Y_{lJ}^* (\hat{x}') Y_{lJ}(\hat{x}) &=& \frac{\textrm{dim} (R_l^{d-1})}
{\textrm{vol}(S^{d-1})} \frac{C_l^\lambda(\mu)}{C_l^\lambda(1)},
\label{eq:ylmclcoeff}
\ee
where $\textrm{vol}(S^{d-1})$ is the volume of the
$(d-1)$-sphere, and 
\be
\textrm{dim}(R_l^{d-1}) &=& {l+d-1 \choose d-1} - {l+d-3 \choose d-1} =
   \frac{d+2l-2}{d-2} { d+l -3 \choose l}
\ee
 is the dimension of the spin-$l$ representation of $SO(d)$.
This allows us to  
simplify the sum over Laurent coefficients:
\be
\< \CO(\hat{x}' ,\tau) \CO(\hat{x}, 0 ) \> &=&
  \frac{1}{\textrm{vol}(S^{d-1}) } \sum_{n,l} N_{n,l}^2(\Delta) e^{-\tau(\Delta+2n+l)}
\frac{d+2l-2}{d-2} C_l^\lambda(\cos \gamma) .
\label{eq:OO1}
\ee
Now, we may also use the explicit form of the 2-point correlator:
\be
\< \CO(\hat{x}', \tau) \CO(\hat{x}, 0) \> &=&
  \frac{e^{\tau \Delta}}{|x-x'|^{2\Delta}} = \frac{1}{2^\Delta (z
 - \mu)^\Delta}, \ \ \ \ \  
z= \cosh \tau, \ \ \  \mu = \cos \gamma .
\label{eq:OO2}
\ee
Eqs. (\ref{eq:OO1}) and (\ref{eq:OO2}) provide two different expressions
for $\< \CO' \CO \>$.  Integrating them against
$(1-\mu^2)^\lambda d \mu C_l^\lambda(\mu)$ and setting them equal, we obtain
\be
&& \frac{1}{\textrm{vol}(S^{d-1})}  \sum_{n,l} N_{n,l}^2(\Delta) e^{-\tau(\Delta+2n+l)}
  \frac{2^{-2\lambda} }{ \lambda \Gamma^2(\lambda) l!} (2\pi) 
\Gamma(2\lambda+ l) \\
&& \ \ \ \ \ \ \ \ \ \ \ \ = \frac{1}{\Gamma(2\lambda)} 
2^{2\lambda +2l} \frac{\Gamma(\lambda + \half) \Gamma(\lambda + \half +l)
\Gamma(\Delta +l)}{\Gamma(\Delta)\Gamma(2\lambda + 2l +1)} e^{-(\Delta +l)\tau}
  F( \Delta +l , \Delta - \lambda, \lambda + l +1, e^{-2\tau}) . \nn
\ee
Expanding the hypergeometric function as a series and matchin powers of
$e^{-\tau}$, we may read off $N_{n,l}^2(\Delta)$:
\be
\frac{1}{N_{n,l}^2(\Delta)}
&=& \textrm{vol} (S^{d-1}) \frac{ \Gamma(\Delta + n +l) \Gamma(\Delta + n -\lambda)
\Gamma(\lambda+1)}{ (n!) \Gamma(\Delta) \Gamma(\Delta - \lambda)
\Gamma(\lambda + n +l + 1)} ,
\ee
which agrees with eq (\ref{eq:Onorm}) in the text.

\bibliography{CFTSMatrixDraft}
\bibliographystyle{JHEP}

\end{document}